\documentclass[12pt]{iopart}
\usepackage{iopams}  
\usepackage{epsfig}
\usepackage{float}
\usepackage{wrapfig}
\usepackage{rotating}
\eqnobysec
\def\be{\begin{equation}}
\def\ee{\end{equation}}
\def\bea{\begin{eqnarray}}
\def\eea{\end{eqnarray}}

\def\sigts{\tilde{\Sigma}_{\sigma}}

\def\sigtdr{\tilde{\Sigma}_{\downarrow}^{\rm R}}
\def\sigtur{\tilde{\Sigma}_{\uparrow}^{\rm R}}
\def\sigtsr{\tilde{\Sigma}_{\sigma}^{\rm R}}
\def\sigtsi{\tilde{\Sigma}_{\sigma}^{\rm I}}
\def\sigs{\Sigma_{\sigma}}
\def\sigu{\Sigma_{\uparrow}}
\def\sigd{\Sigma_{\downarrow}}
\def\sigur{\Sigma^{\rm R}_{\uparrow}}
\def\sigdr{\Sigma^{\rm R}_{\downarrow}}
\def\sigui{\Sigma^{\rm I}_{\uparrow}}
\def\sigdi{\Sigma^{\rm I}_{\downarrow}}
\def\sigsr{\Sigma^{\rm R}_{\sigma}}
\def\sigsi{\Sigma^{\rm I}_{\sigma}}
\def\w{\omega}
\def\sgn{{\rm sgn}}

\def\gmb{g\mu_{\rm B}}
\def\sg{{\cal G}}
\def\da{\downarrow}
\def\ua{\uparrow}

\def\ra{\rightarrow}
\def\ppm{\Pi^{+-}}
\def\pnpm{{^0\Pi^{+-}}}
\def\eg{{\em e.g. }}
\def\ie{{\em i.e. }}
\def\ut{\tilde{U}}
\def\delno{\Delta_0}
\def\pdod{\pi\Delta_0 D}
\def\w{\omega}
\def\wm{{\omega_{\rm m}}}
\def\wmo{{\omega_{\rm m}^0}}
\def\wmh{{\omega_{\rm m}(h)}}
\def\wmph{{\omega^\prime_{\rm m}(h)}}
\def\wp{\omega^\prime}
\def\wmp{{\omega_{\rm m}^\prime}}
\def\wk{{\omega_{\rm K}}}
\def\tk{{T_{\rm K}}}
\def\wa{{\omega_{\alpha}}}
\def\wt{{\tilde{\omega}}}

\def\im{{\rm Im}}

\def\bra{\langle}
\def\ket{\rangle}

\def\PRB{{\em Phys. Rev. B }}

\def\ut{\tilde{U}}
\def\ofwh{(\omega;h)}
\def\ofoh{(0;h)}
\def\ofwo{(\w;0)}
\def\hp{h^\prime}
\def\htilde{\tilde{h}}
\def\zh{Z(h)}
\def\zo{Z(0)}
\def\rw{R_{\rm W}}
\def\wct{\tilde{\w}_{\rm c}}
\def\wpeaks{\w_{\rm p}}

\begin{document}

\title{Field-dependent dynamics of the Anderson impurity model.}

\author{David E. Logan and Nigel L. Dickens }

\address{University of Oxford, Physical and Theoretical Chemistry\\ Laboratory, South Parks Rd, Oxford OX1 3QZ, UK}

\begin{abstract}
Single-particle dynamics of the Anderson impurity model in the presence of a magnetic field $H$ are considered, using a recently developed local moment approach that encompasses all energy scales, field and interaction strengths. For strong coupling in particular, the Kondo scaling regime is recovered. Here the frequency ($\omega/\wk$) and field ($H/\wk$) dependence  of the resultant universal scaling spectrum is obtained in large part analytically, and the field-induced destruction of the Kondo resonance investigated.  The scaling spectrum is found to exhibit the slow logarithmic tails recently shown to dominate the zero-field scaling spectrum. At the opposite extreme of the Fermi level, it gives asymptotically exact agreement with results for statics known from the Bethe ansatz. Good agreement is also found with the frequency  and field-dependence of recent numerical renormalization group calculations. Differential conductance experiments on quantum dots in the presence of a magnetic field are likewise considered; and appear to be well accounted for by the theory. Some new exact results for the problem are also established.
\end{abstract}

\pacs{71.27.+a, 72.15.Qm, 75.20.Hr}

\submitto{\JPCM}

%\maketitle

\section{Introduction}
\label{sec:1}

The Anderson impurity model (AIM) \cite{ref:anderson} has long occupied a central role in condensed matter theory.  Reviewed comprehensively in \cite{ref:hewson}, it serves as a paradigm for the physics of strong local interactions, and remains the canonical model for understanding magnetic impurities in metals; competition between on-site Coulomb repulsion and band hybridization generating the Kondo effect in strong coupling, where the former dominates the latter.  Renewed interest in the problem has arisen recently from the discovery that direct mesoscopic realizations of AIMs may be tailor made: quantum dots \cite{ref:qdots,ref:cok} for example, or surface atoms probed by scanning tunneling microscopy \cite{ref:stm}.  Moreover the intrinsic tunability of such nanoscale devices offers controlled access to a wider range of quantum `impurity' physics than usually accessible with more traditional materials.

Experimental probes of quantum dots are of course typically dynamical; and therein lies a major theoretical challenge.  Static (thermodynamic and related) properties of the AIM are well understood, using a variety of powerful techniques that include the numerical renormalization group (NRG) \cite{ref:NRG}, Fermi liquid theory \cite{ref:noz}, and the Bethe ansatz \cite{ref:tw}.  That is not however the case for dynamical properties, such as the single-particle excitations.  The many theoretical approaches to such are approximate by necessity, and suffer from well known qualitative limitations \cite{ref:hewson} -- even for the AIM in equilibrium, let alone non-equilibrium effects that, to a greater or lesser extent, are relevant to \eg non-linear differential conductance measurements \cite{ref:qdots,ref:cok}.  And the fact that  numerical methods, notably the NRG, can now provide benchmark numerical results for equilibrium dynamics (see \eg \cite{ref:hewson}) gives added impetus to further development of approximate many-body theories.

%hard cite
In this paper we consider the AIM in the presence of an applied magnetic field, $H$.  This is a topical issue [9-15], motivated in part by differential conductance measurements on quantum dots in the Kondo regime \cite{ref:qdots,ref:cok}:  these show a characteristic field-induced splitting of the Kondo resonance \cite{ref:mwl}, indicative of the slow crossover from a locked Kondo singlet to an asymptotically free local moment with increasing $h = \frac{1}{2} \gmb H$.  We pursue it here within the framework of the recently developed local moment approach (LMA) [16-19], a technically straightforward non-perturbative many-body method in which the notion of local moments \cite{ref:anderson} is introduced explicitly and self-consistently from the outset.

The LMA handles single-particle dynamics on all energy scales, and for all interaction strengths $\tilde{U} = U / \pi \delno$ \cite{ref:L1,ref:L2} (with $U$ the on-site Coulomb repulsion and $\delno$ the hybridization strength).  Most importantly, in strong coupling $\tilde{U} \gg 1$ it captures the Kondo or spin fluctuation regime characterized by a low-energy Kondo scale $\wk$ that is exponentially small, such that zero-field single particle dynamics exhibit universal scaling in terms of $\w / \wk$.  The resultant scaling spectrum for the symmetric AIM yields good quantitative agreement with NRG results \cite{ref:bulla1,ref:dl}; recovering not only Fermi liquid behaviour on the lowest energy scales, but also revealing slow logarithmic tails that in fact dominate the scaling spectrum \cite{ref:dl}.

For $h \ne 0$, LMA results for static properties have recently been considered \cite{ref:ld}, notably for the impurity magnetization and corresponding $h$-dependent spin susceptibility in the Kondo regime.  These too are found to yield good agreement with exact results known from the Bethe ansatz \cite{ref:tw,ref:afl}, being asymptotically exact in both the weak- and strong-field limits; and correctly recovering the field-independence of the Wilson ratio, $\rw (h) = 2$ for all $h$ \cite{ref:tw,ref:wf}.  Here by contrast we consider field-dependent spectral dynamics, beginning  (\S 2) with a brief introduction to the LMA for finite $h$; and focussing on the symmetric model, in which the spectral effects of an applied field are most simply apparent.  There are two essential domains of field strength.  First the important strong coupling, universal Kondo limit, where $h$  by definition is irrelevantly small compared to the `bare' electronic scales $\delno$ or $U$; but with $h / \wk$ arbitrary, and spanning the entire range of field strengths relevant to the Kondo model.  It is this regime that is of primary interest  in relation to experiments on quantum dots \cite{ref:qdots,ref:cok}.  There are also non-universal regimes of field strength, including a striking spectral signature of the crossover to the high-field limit where the fermions become effectively spinless.  These are discussed as part of \S 3, where evolution of the AIM spectra on all frequency and field scales is considered, with the aim of extracting a broad overall picture of the problem.

In \S 's 4,5 we turn to the strong coupling Kondo limit, and spectral scaling in terms of $\w / \wk$ and $h / \wk$.  As for the zero-field case considered recently \cite{ref:dl}, our initial aim (\S 4) is to obtain analytically the $h$-dependent scaling spectrum in a manner that is largely independent of the details of the LMA; and in particular to deduce explicitly (\S 4.1) the behaviour of the high-frequency spectral tails, as well as the field-dependence of the spectrum at the Fermi level.  The main body of results for the LMA scaling spectra are given in \S 5; including comparison to recent NRG \cite{ref:costi} and density matrix NRG (DM-NRG) \cite{ref:hof,ref:hofpriv} calculations, as well as to the spinon approximation in which \cite{ref:mw} the single-particle spectrum is approximated by the density of states for spinon excitations obtained from the Bethe ansatz.  The LMA results of \S 5.2 are found to compare favourably with the former, but are markedly at odds with those of the spinon approximation.  The latter is discussed explicitly in \S 5.3, and its qualitative limitations identified.

Finally, comparison is made in \S 5.4 to recent differential conductance experiments \cite{ref:qdots} on quantum dots in a magnetic field; which, notwithstanding the natural absence of non-equilibrium  effects in the LMA, are found to be rather well described by the theory.  A brief summary is given (\S 6), together with an appendix in which, using microscopic Fermi liquid theory, we obtain two exact results for the problem that to our knowledge are new; specifically for the field-dependence of the quasiparticle weight, and the asymptotic low-field behaviour of the spectral shifts.

\section{Background}
\label{sec:2}

The Hamiltonian for the AIM \cite{ref:anderson} is given in conventional notation by

\be
\label{eq:ham}
\hat{H} = \sum\limits_{\bi{k}, \sigma} \epsilon_{\bi{k}} \hat{n}_{\bi{k}\sigma} + \sum\limits_{\sigma} \left( \epsilon_{i\sigma} + \frac{U}{2} \hat{n}_{i-\sigma} \right) \hat{n}_{i\sigma} + \sum\limits_{\bi{k}, \sigma} V_{i\bi{k}} \left( c_{i\sigma}^\dagger c_{\bi{k}\sigma} + {\rm h.c.} \right).
\ee

\noindent The first term refers to the host band of non-interacting electrons, with dispersion $\epsilon_\bi{k}$.  The second refers to the impurity, with on-site interaction $U$ and site-energy $\epsilon_{i\sigma} = \epsilon_{i}-\sigma h$.
The latter includes a local Zeeman coupling to the external field $H$ (applied for convenience in the $-z$-direction), with $h=\frac{1}{2}\gmb H$ and $\sigma = +/-$ for $\ua/\da$-spin electrons.
The final term in equation~(\ref{eq:ham}) is the one-electron host-impurity coupling.  For the symmetric AIM that we consider, $\epsilon_i = -\frac{U}{2}$, and by particle-hole (p-h) symmetry $n_i = \sum\nolimits_\sigma \langle \hat{n}_{i\sigma} \rangle = 1$ for all $U$ and $h$.

We focus on the total impurity Green function $G(\w; h)$ (with corresponding spectral density $D\ofwh = -\pi^{-1} \sgn(\w)~\im G\ofwh $); where

\be
\label{eq:g_to_gsig}
G\ofwh  = \frac{1}{2}\sum\limits_\sigma G_\sigma \ofwh 
\ee

\noindent with $G_\sigma \ofwh  = G^{\rm R}_\sigma \ofwh  - \rmi~\sgn(\w)\pi D_\sigma \ofwh $, given by

\be
\label{eq:gsig}
G_\sigma(\w,h) = \left[ \w^+ - \Delta (\w) + \sigma h - \sigts \ofwh  \right]^{-1}
\ee

\noindent and $\w^+ = \w + \rmi 0^+\sgn (\w)$.  Here $\Delta(\w) = \Delta_{\rm R}(\w) - \rmi~\sgn(\w)\Delta_{\rm I}(\w)$ ($= -\Delta(-\w)$) is the host-impurity hybridization, with $\Delta_{\rm I} (\w) = \pi \sum_{\bi{k}} V_{i\bi{k}}^2 \delta (\w - \epsilon_{\bi{k}})  $; and the hybridization strength $\delno = \Delta_{\rm I}(\w = 0)$ is thus defined, with $\w = 0$ the Fermi level.
$\sigts \ofwh  = \sigtsr\ofwh - \rmi~\sgn(\w)\sigtsi \ofwh $ denotes the impurity self-energy (excluding the trivial Hartree contribution that precisely cancels $\epsilon_i = -\frac{U}{2}$).  By p-h symmetry

\be
\label{eq:sigsym}
\sigts \ofwh  = -\tilde{\Sigma}_{-\sigma}(-\w;h)
\ee

\noindent and likewise $D_\sigma\ofwh  = D_{-\sigma}(-\w;h)$; whence $D\ofwh  = \frac{1}{2} \sum_\sigma D_\sigma\ofwh  = D(-\w;h)$ is naturally symmetric in $\w$ about the Fermi level.  The $h$-dependent quasiparticle weight $Z(h)$ is defined by

\be
\label{eq:zhdef}
Z(h) = \left[ 1 - ( \partial \sigtsr\ofwh  / \partial \w )_{\w = 0} \right] ^{-1}
\ee

\noindent (and from equation~(\ref{eq:sigsym}) is $\sigma$-independent).

In practice we consider explicitly the usual wide-band AIM for which $\Delta_{\rm I}(\w) = \delno ~ \forall~ \w$, and $\Delta_{\rm R} (\w) = 0$.  This is not of course restrictive since in strong coupling, $\ut = U/\pi\delno \gg 1$, the relevant low-energy Kondo resonance is a universal function of $\w / \wk$ (with the Kondo scale $\wk$ defined as the HWHM of the $h=0$ Kondo resonance).  It is thus independent of the detailed one-electron structure of $\Delta(\w)/\delno$, which affects only the dependence of $\wk$ itself on the bare one-electron parameters (such as the bandwidth, D, of $\Delta_{\rm I}(\w)$).
By the same token the $h$-dependence of the Kondo resonance is independent of whether a local or global magnetic field is considered.  Application of a global uniform field leads additionally to an $h$-dependent hybridization $\Delta_\sigma \ofwh  = \Delta(\w + \sigma h)$; but its $h$-dependence arises on a one-electron energy scale, and is thus irrelevant in the Kondo regime of finite $h / \wk$ and $\wk \propto \exp (-\pi U / 8 \delno) \ra 0$.

%hard ref
Within the LMA [16-19] the self-energy $\sigts \ofwh $  is separated as

\be
\label{eq:sig_to_sigt}
\sigts\ofwh  = -\frac{\sigma}{2} U |\mu(h)| + \sigs \ofwh 
\ee

%\begin{wrapfigure}{l}{75mm}
\begin{figure}
\centering\epsfig{file=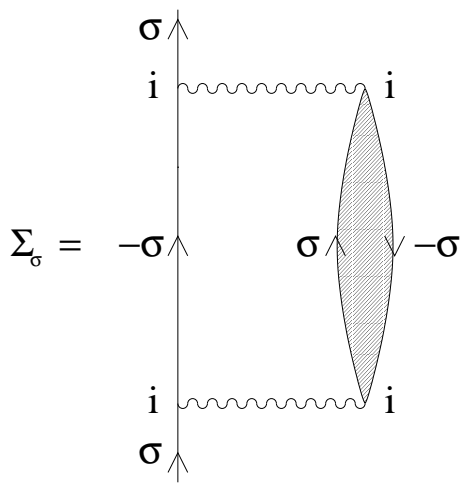,width=60mm,angle=0}
%%% \vskip-1mm
\protect\caption{Principal contribution to the LMA $\Sigma_\sigma(\omega)$, see text.  Wavy lines denote $U$.}
\label{fig:diag}
\end{figure}
%\end{wrapfigure}

\noindent into a static Fock contribution (with local moment $|\mu(h)|$) that alone would survive at the simple mean-field (MF) level of unrestricted Hartree-Fock; together with a dynamical contribution $\sigs\ofwh $.  The latter includes in particular a non-perturbative class of diagrams (figure~\ref{fig:diag}) that embody dynamical coupling of single-particle excitations to low-energy transverse spin fluctuations.  Other classes of diagrams may also be included, but retention of the dynamical spin-flip scattering processes is essential to capture the strong coupling Kondo regime for $\ut \gg 1$ \cite{ref:L1,ref:L2}.  These are expressed in terms of MF propagators (solid lines in figure~\ref{fig:diag}), viz

\be
\label{eq:scriptgdef}
\sg_\sigma \ofwh  = \left[ \w^+ - \Delta(\w) + \sigma\left(\frac{1}{2} U |\mu(h)| + h\right) \right]^{-1}
\ee

\noindent with spectral densities $D_\sigma^0 (\w) = -\pi^{-1}\sgn(\w)~\im \sg_\sigma \ofwh $; and $\sigu \ofwh $ ($ = -\sigd (-\w;h)$) is given explicitly by \cite{ref:L1}

\be
\label{eq:sigint}
\hspace{-1.5cm}\sigu\ofwh  \! = \! U^2 \!\!\! \int\limits^\infty_{-\infty} \frac{\rmd \w_1}{2\pi\rmi} \im \ppm \ofwh  \left[ \theta(\w_1) \sg_\da^- (\w_1 \! + \! \w;h) + \theta(-\w_1) \sg_\da^+ (\w_1 \!+ \!\w;h) \right]
\ee

\noindent with $\sg_\sigma^\pm \ofwh $ the one-sided Hilbert transforms of $\sg_\sigma \ofwh $ and $\theta(x)$ the unit step function.  Here, $\ppm\ofwh$ denotes the transverse spin polarization propagator (shown hatched in figure~\ref{fig:diag}).  It is given at the simplest level by an RPA-like particle-hole ladder sum in the transverse spin channel, \ie $\ppm = \pnpm / (1 - U\pnpm)$ with $\pnpm \ofwh$ the bare p-h bubble,  itself expressed in terms of MF propagators.

The central physical idea behind the LMA for $h=0$ is that of symmetry restoration, as detailed in \cite{ref:L1,ref:L2} : restoration of the broken symmetry endemic at pure MF level,  via the spin-flip dynamics embodied in $\sigs (\w;0)$.  The MF propagators form the basis for constructing the dynamical self-energies $\sigs \equiv \sigs [\sg_\sigma]$ shown in figure~\ref{fig:diag};  and by themselves correspond to local symmetry breaking for $|\mu(0)| > 0$.  Symmetry restoration is embodied mathematically in $\sigtur (0;0) = \sigtdr (0;0)$, \ie using p-h symmetry by

\be
\label{eq:sigpin}
\sigtur (0;0) = -\frac{1}{2} U |\mu(0)| + \sigur (0;0) = 0.
\ee

\noindent Equation~(\ref{eq:sigpin}) is imposed self-consistently, achieved in practice \cite{ref:L1,ref:L2} for given $\ut$ by varying the local moment $|\mu(0)|$ from its pure mean-field value $|\mu_0 |$ (where $\mu_0 = \bra \hat{n}_{i\ua} - \hat{n}_{i\da} \ket_0$ with the average over the MF ground state).
Symmetry restoration ensures correctly that the low-$\w$ behaviour of $G(\w;0)$ constitutes a renormalization of the non-interacting limit; as manifest \cite{ref:L1} in a resultant quasiparticle form for $G(\w;0)$, the recovery of Fermi liquid behaviour at low energies, and preservation of the $U$-independent pinning of the Fermi level spectrum (viz $\pdod (0;0) = 1~\forall~ U \ge 0$, as follows directly from equations~(\ref{eq:g_to_gsig}, \ref{eq:gsig}) using equation~(\ref{eq:sigpin})).

Most importantly, self-consistent imposition of equation~(\ref{eq:sigpin}) introduces naturally a low-energy spin-flip scale $\wmo = \w_{\rm m} (h=0)$, evident in particular in a strong resonance centred on $\w = \wmo$ in the transverse spin polarization propagator $\im\ppm(\w;0)$.  Its form in strong coupling is readily deduced from equation~(\ref{eq:sigint}) for $\w = 0 = h$, which for $\ut \gg 1$ has the asymptotic form \cite{ref:L1}

\be
\label{eq:scsigr}
\sigur(0;0) = \frac{4\delno}{\pi}  \ln \left[ \frac{\lambda}{\wmo} \right]
\ee

\noindent with $\lambda = {\rm min}[\frac{U}{2}, D]$.  Combining equation~(\ref{eq:scsigr}) with the symmetry restoration condition equation~(\ref{eq:sigpin}) (using $|\mu(0)| \ra 1$ in strong coupling) gives $\wmo \sim \lambda \exp (-\pi U / 8 \delno)$.  This is the Kondo scale,  exponentially small in strong coupling and recovering the exact exponent \cite{ref:hewson}.  It naturally has no counterpart at simple MF level, and within the LMA the physical significance of the Kondo scale $\wmo$ ($\propto \wk$) is that it sets the timescale ($\sim \hbar / \wmo$) for restoration of the broken symmetry inherent at pure MF level.

Before turning to $h \ne 0$, note that for $h=0$ two degenerate MF states arise \cite{ref:L1}, reflecting simply the invariance of $\hat{H}$ under $\sigma \ra -\sigma$ ($\ua / \da$-spin symmetry).  These states, denoted temporarily by $\alpha = {\rm A}$ or ${\rm B}$, correspond respectively to a local moment of $\mu = \pm |\mu|$.  The corresponding MF propagators are then denoted in full by $\sg_{\alpha\sigma} (\w;0)$, the self-energies constructed from them by $\tilde{\Sigma}_{\alpha\sigma} (\w;0)$, and the resultant many-body Green functions by $G_{\alpha\sigma} (\w;0)$; for example, equation~(\ref{eq:g_to_gsig}) in full notation is $G(\w;0) = \frac{1}{2}\sum\limits_\sigma G_{\alpha\sigma} (\w;0)$ with the $G_{\alpha\sigma}(\w;0)$ given from equation~(\ref{eq:gsig}) in terms of $\tilde{\Sigma}_{\alpha\sigma} (\w;0)$.  In writing equations~(\ref{eq:g_to_gsig}-\ref{eq:zhdef}) we have assumed implicitly that either of the two mean-field states may be employed for $h=0$.  This is indeed correct, for the $G_{\alpha\sigma}(\w;0)$'s are related by $\ua/\da$-spin symmetry, viz

\be
\label{eq:gab}
G_{{\rm A}\sigma}(\w;0) = G_{{\rm B}-\sigma}(\w;0).
\ee

\noindent Hence

\be
\label{eq:gasum}
G(\w;0) = \frac{1}{2} \sum\limits_\sigma G_{\alpha\sigma} (\w;0)
\ee

\noindent is independent of $\alpha$ (which is not moreover specific to the symmetric AIM, since p-h symmetry has not been used in any way).  Using equation~(\ref{eq:gab}), $G(\w;0)$ may also be written equivalently as

\be
\label{eq:absum}
G(\w;0) = \frac{1}{2} \sum\limits_\alpha G_{\alpha\sigma} (\w;0).
\ee

%hard ref
\noindent This form shows that $G(\w;0)$ may be viewed equivalently as involving an average over the two degenerate MF states $\alpha =  A$, $B$. Equation~(\ref{eq:absum}) is moreover independent of spin, $\sigma$, reflecting the fact that $G(\w; 0)$ is equivalently the $\sigma$-spin Green function. Hence $G(\w;0) = [\w^+ - \Delta(\w) - \Sigma(\w;0)]^{-1}$ in terms of the conventional single self-energy $\Sigma(\w;0)$ which, using equation~(\ref{eq:absum}), may therefore be obtained from the underlying two-self-energy description inherent to the LMA as discussed in [16-18].

The above situation naturally changes for $h \ne 0$.  The degeneracy is removed and one or other MF state is picked out, according to $\sgn (h)$.  $G\ofwh $ is then given by

\be
\label{eq:gasumh}
G\ofwh  = \frac{1}{2} \sum\limits_\sigma G_{\alpha\sigma} \ofwh 
\ee

\noindent with $\alpha = {\rm A}$ for $h > 0$ and B for $h < 0$ (given our convention for the Zeeman coupling in the Hamiltonian, equation~\ref{eq:ham}).  But the invariance of $\hat{H}$ under $(\sigma, h) \ra (-\sigma,-h)$ implies

\be
\label{eq:habsym}
G_{{\rm A}\sigma} (\w;|h|) = G_{{\rm B}-\sigma} (\w;-|h|)
\ee

\noindent  (and likewise for the $\tilde{\Sigma}_{\alpha\sigma}\ofwh$), and thus $G\ofwh  = G(\w;-h)$.  Only $h>0$ need therefore be considered, and hence $\alpha = {\rm A}$;  this will be assumed henceforth (and is already implicit in equations~(\ref{eq:g_to_gsig}-\ref{eq:scsigr})), and the $\alpha$ label thus dropped.  Finally, note that while $G\ofwh $ evolves continuously in $h$ to its $h=0$ limit $G(\w;0)$, the fact that the degeneracy of the MF states is removed for $h \ne 0$ means that $G\ofwh $ for $h \ne 0$ is naturally not expressible in the form of equation~(\ref{eq:absum}).

\section{Dynamics: all scales.}
\label{sec:3}

We first consider single-particle dynamics on all frequency and field scales, \ie encompassing both the low-energy Kondo resonance and high-energy Hubbard satellites, as well as the full range of magnetic field strengths.  The $h$-dependence of the low-energy scaling spectrum that arises in the Kondo limit will be pursued in the following sections.

$G\ofwh $ is given by equations~(\ref{eq:g_to_gsig}), and using equations~(\ref{eq:gsig}, \ref{eq:sig_to_sigt}, \ref{eq:scriptgdef}) $G_\sigma\ofwh $ may be expressed as $G_\sigma\ofwh  = [\sg_\sigma^{-1}\ofwh  - \sigs\ofwh ]^{-1}$ in terms off the dynamical part of the self-energy, $\sigs\ofwh$.  The latter, given in practice by equation~(\ref{eq:sigint}) and figure~\ref{fig:diag}, is quite generally a functional of the MF propagators $\sg_\sigma \ofwh$ given by equation~(\ref{eq:scriptgdef}):  $\sigs \ofwh \equiv \sigs [\sg_\sigma]$.
Hence $G_\sigma \ofwh \equiv G_\sigma [\sg_\sigma]$, and from equation~(\ref{eq:scriptgdef}) it follows that the $h$-dependence is embodied fully in $x(h) = \frac{1}{2} U |\mu(h)| + h$.  The LMA $G \ofwh$ for $h > 0$ is thus formally equivalent to that for $h=0$, but with $x(0)$ replaced by $x(h) = x(0) + [\frac{1}{2} U \delta|\mu(h)| +h]$; where $\delta|\mu(h)| = |\mu(h)| - |\mu(0)|$, and the zero-field moment $|\mu(0)|$ is determined such that symmetry restoration (equation~(\ref{eq:sigpin})) is satisfied.

In practice we take $\delta|\mu(h)|$ to be given approximately at MF level, specifically for the wide-band AIM by

\be
\label{eq:dmu}
\delta|\mu(h)| \simeq \frac{2}{\pi} \left\{ \tan^{-1}\left( \frac{\frac{1}{2}U|\mu(0)|+h}{\delno}\right) - \tan^{-1} \left( \frac{\frac{1}{2}U|\mu(0)|}{\delno}\right) \right\}.
\ee

\noindent The $h$-dependence of $\delta|\mu(h)|$ is in fact of little importance, since it is never significant compared to the bare Zeeman term $(h)$ in $x(h) - x(0) = \frac{1}{2} U \delta|\mu(h)| +h$.  In strong coupling (SC) $\ut = U/\pi\delno \gg 1$, this is seen directly from equation~(\ref{eq:dmu}) which yields

\be
\label{eq:scdmu}
\frac{U}{2} \delta|\mu(h)| \sim \left[ \frac{2}{\pi} \right]^2 \frac{1}{\ut} \frac{h}{(1+h / [U/2])}
\ee

\noindent (where we use $|\mu(0)| \ra 1$ in SC \cite{ref:L1}, reflecting saturation of the moment).  And in weak coupling $\ut \ll 1$ (where $|\mu(0)| = 0$ \cite{ref:L1}), $\frac{1}{2}U\delta|\mu(h)| / \delno \sim \ut \tan^{-1} (h/\delno)$ is likewise insignificant compared to the Zeeman term $h/\delno$; although its retention is readily shown to be required to recover exactly the leading $\ut$-dependence of the static impurity susceptibility $\chi_\rmi (0)$ (LMA results for which have been considered in \cite{ref:ld}).

LMA spectra for arbitrary $h$ now follow straightforwardly.  The zero-field $x(0) = \frac{1}{2} U |\mu(0)|$ is first determined using the $\w = 0$ symmetry restoration condition, equation~(\ref{eq:sigpin}); and $x(h)$ for any $h > 0$ follows as above.  $\sigu\ofwh ( = - \sigd(-\w;h))$ for all $\w$ is given by equation~(\ref{eq:sigint}) with the MF propagators from equation~(\ref{eq:scriptgdef}).  $G_\sigma \ofwh = [\sg_\sigma^{-1} - \sigs ]^{-1}$, $G \ofwh = \frac{1}{2}\sum\nolimits_\sigma G_\sigma \ofwh$ and the single-particle spectrum $D\ofwh$ then follows directly.

Before proceeding we comment briefly on the $h$-dependence of the spin-flip scale $\wmh$, defined (as for $h=0$ \cite{ref:L1,ref:L2}) as the position of the maximum in the transverse spin polarization propagator $\im \ppm \ofwh$; itself given (\S 2) by the p-h ladder sum, with the bare polarization bubble $\pnpm \ofwh$ expressed in terms of MF propagators.  But since $\wmh \equiv \wm (x(h))$ as above, the $h$-dependence of the spin-flip scale follows from a knowledge of the $x$-dependence of $\pnpm$ as considered in \cite{ref:L1}.  This may be deduced analytically in SC, with the result

\be
\label{eq:scwmh}
\wmh = \wmo + 2 |\mu_0|h.
\ee

\noindent Here $|\mu_0|$ is the MF local moment in zero-field, given explicitly from $|\mu_0| = \frac{2}{\pi}\tan^{-1}\left( \frac{\pi}{2} \ut |\mu_0| \right)$ for the wide-band AIM \cite{ref:L1}; and in practice equation~(\ref{eq:scwmh}) is numerically accurate for $\ut \gtrsim 3$ or so.  In the SC Kondo limit where $|\mu_0| \ra 1$, $\wmh = \wmo + 2h \sim 2h$ for $h / \wmo \gg 1$.  This is physically correct, it being known from solution of the Kondo/s-d model \cite{ref:tw,ref:afl} that for fields large compared to the Kondo scale $\wk \sim \wmo$ -- and in practice for $h / \wk \gtrsim 1-10$ -- the impurity spin becomes asymptotically free (albeit with logarithmic corrections to the impurity magnetization $M_\rmi (h)$ \cite{ref:tw,ref:afl}); and for a free spin-$\frac{1}{2}$ the sole energy scale for spin flips is the Zeeman splitting $2h = \epsilon_{i\ua} - \epsilon_{i\da}$ characteristic of the atomic limit.

%hard ref
Single-particle spectra for $h=0$ have been considered in references~[16-18] to which the reader is referred (and are shown as part of figures~2, 5, 7 below).  While the LMA is perturbatively exact to second order in $U$ in weak coupling \cite{ref:L1}, our primary interest is naturally in SC where the Kondo effect prevails.  Figure~$2a$ thus shows the zero-field spectrum \cite{ref:L1} (solid line) $\pdod (\w;0)$ vs $\w / \delno$ for $\ut = 6$ (wide-band AIM).  The following spectral features should be noted.  (i) The Kondo resonance is correctly pinned at the Fermi level, $\pdod (0;0) = 1$, as follows from self-consistent imposition of symmetry restoration (equation~(\ref{eq:sigpin})).  By the same token its HWHM, the Kondo scale $\wk$, is exponentially small: $\wk \propto \wmo$ with $\wmo \propto \exp(-\pi U / 8 \delno)$ in SC as explained in \S 2.  (ii) The maximum in the Hubbard satellite(s) occurs at $|\w| = \frac{U}{2}$ in SC.  It is Lorentzian in form, but its HWHM is $2\delno$ and not $\delno$ as simple MF theory would predict, due to additional many-body broadening processes \cite{ref:L1,ref:pg} (see also below).

\subsection{Results: $h > 0$.}
\label{sec:3.1}

There are three energy scales relevant to the $h = \frac{1}{2}\gmb H$ dependence of single-particle spectra, namely the Kondo scale $\wmo$, the hybridization $\delno$, and the interaction $\frac{U}{2}$; and such that $\wmo \ll \delno \ll U$ in SC.  There are thus several relevant domains of field strength.  First the all important Kondo limit, corresponding formally to finite $h / \wmo$ and $\wmo \ra 0$.  Here by definition $h$ is vanishingly small compared to $\delno$ or $U$; but $h / \wmo$ is arbitrary, and spans the entire range of field strengths appropriate to the Kondo model.  This is the universal Kondo scaling regime; we consider it in detail in \S 4ff.  Note however that this regime is simply inaccessible to approximate theories in which the Kondo scale does not exist (\eg equation of motion approaches \cite{ref:mwl}); and to those that fail to recover an exponentially small Kondo scale in SC, and hence the pristine separation between the Kondo scale and $\delno$ or $U$ that is the essence of the Kondo regime (\eg modified perturbation theory \cite{ref:ts}).  It is primarily this domain that is of experimental interest in the context of quantum dots \cite{ref:qdots,ref:cok}.

Figure~2 illustrates spectral evolution in the non-universal regimes of field strength.  For $\ut = 6$, and in addition to $h=0$, figure~$2a$ shows the LMA $\pdod \ofwh$ vs $\w/\delno$ for $h / \wmo = 125$ ($h / \delno \sim  0.15 \ll 1$) and $h / \wmo = 10^3$ ($h \sim \delno$); this is continued in figure~$2b$ for $h / \wmo = 10^3$, $4\times 10^3$ ($\delno < h < \frac{U}{2}$), and $h / \wmo = 1.2\times 10^4$ and $1.6\times 10^4$ illustrating $h > \frac{U}{2}$.

\begin{figure}
\centering\epsfig{file=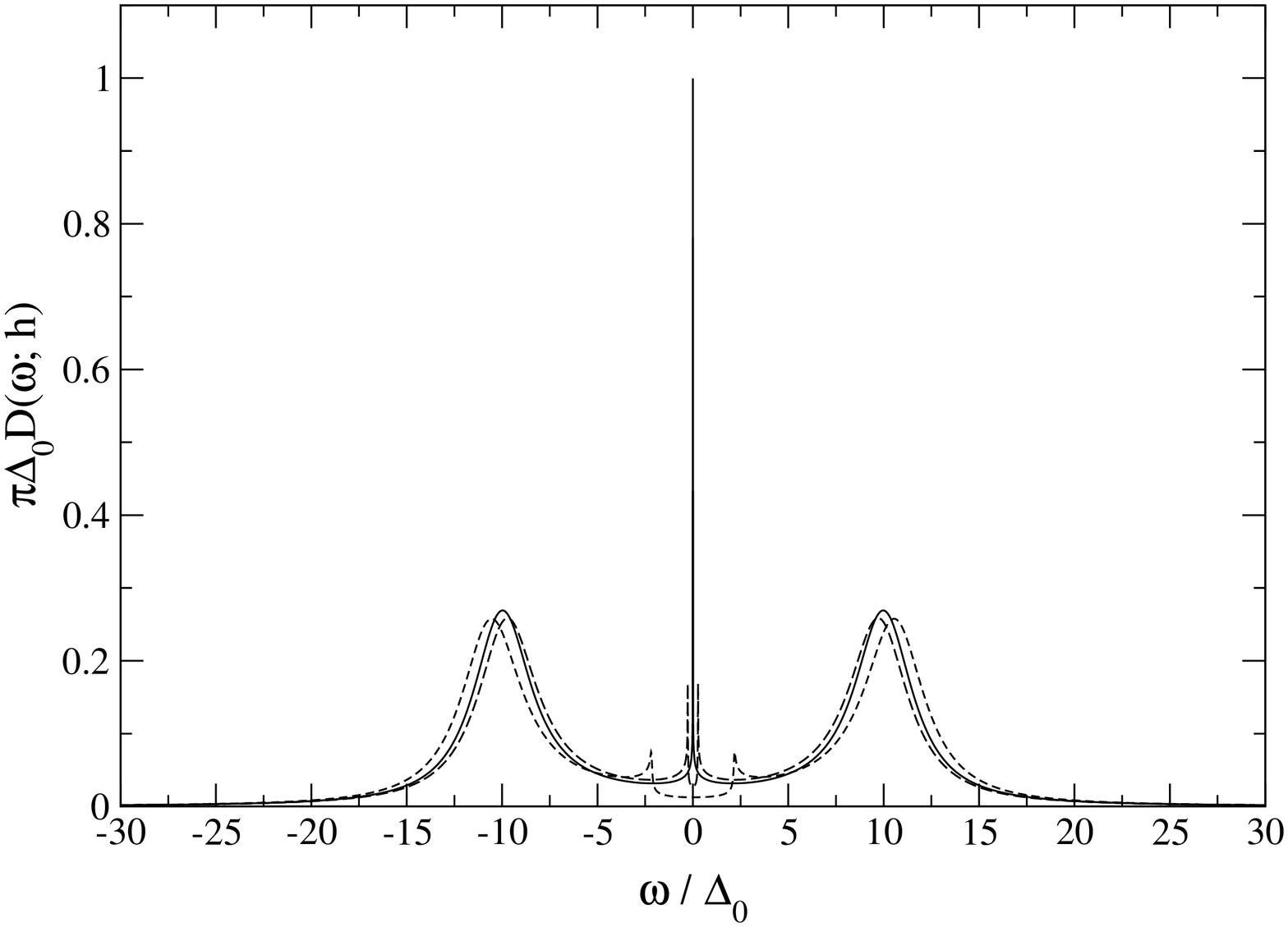,width=100mm,angle=270}
\centering\epsfig{file=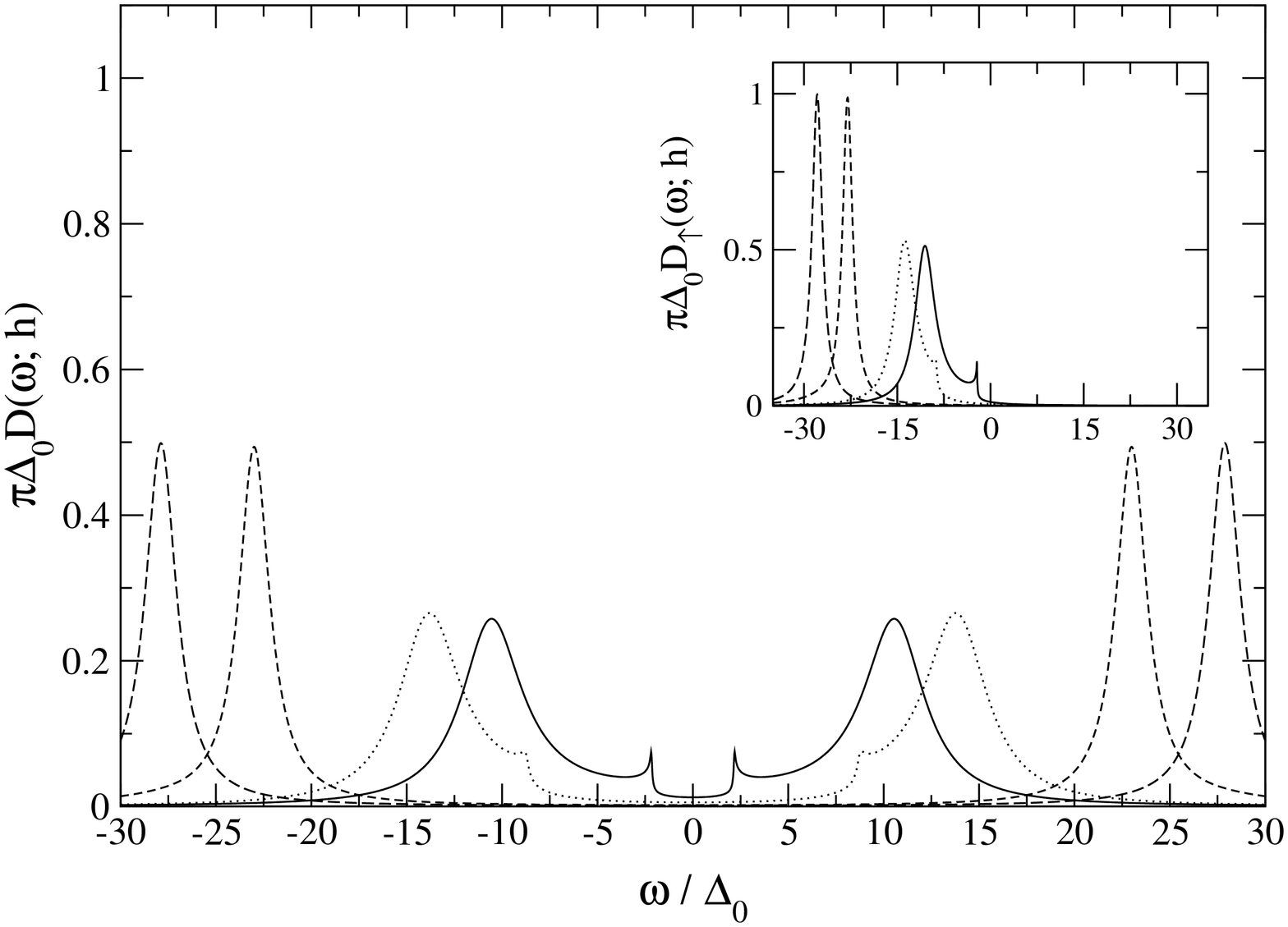,width=100mm,angle=270}
\vskip-5mm
\protect\caption{LMA spectra $\pdod \ofwh$ vs $\w / \delno$ for $\tilde{U} = 6$ as discussed in text.  (a) Upper panel: for $h / \wmo = 0$ (solid line), $125$ (long dash), $10^3$ (short dash).  (b) Lower panel: for $h / \wmo = 10^3$ (solid), $4\times 10^3$ (dotted), $1.2 \times 10^4$ (short dash), $1.6 \times 10^4$ (long dash.  Inset: corresponding $\pdod_\ua \ofwh$ vs $\w / \delno$.}
\label{fig:2}
\end{figure}

Consider first the evolution of the `low'-energy spectral features.  By $h / \wmo = 125$ the zero-field Kondo resonance has split, producing narrow peaks of much reduced intensity centred upon $|\w| \simeq 2h$ \ie the Zeeman splitting energy.  The latter is also found by an equation of motion approach \cite{ref:mwl}, and seen in DM-NRG calculations by Hofstetter \cite{ref:hof} for $\ut \simeq 3$ (see figure~3 of \cite{ref:hof} for $D_\ua \ofwh$, noting that the field therein corresponds to $2h$ in the present notation and that $2h/\delno$ lies in the range $0.1$ to $1$).
We add again however that frequencies of order $|\w| = 2h$ lie within the Kondo scaling `window' only as $2h / \delno \ra 0$, which is not the case for the results shown in figure~2 or those of reference~\cite{ref:hof}.
With further increasing field the split peaks associated with the erstwhile Kondo resonance remain centred on $|\w| \simeq 2h$, but diminish further in intensity; ultimately losing their integrity and being subsumed into the Hubbard satellites as $h$ approaches the order $U/2$ characteristic  of the zero-field satellites.  This is seen further from the LMA $\pdod_\ua \ofwh$ shown in figure~$2b$ inset; it is again qualitatively consistent with the DM-NRG results of reference~\cite{ref:hof}.

There are in fact two non-universal regimes of field strength, the crossover between which occurs for $h \sim \frac{U}{2}$.  This is apparent from the $h$-dependence of the Hubbard satellites.  In SC the latter (figure~2) are centred on $|\w| = \frac{U}{2} + h$, as is obvious from the atomic limit of the model; and the $h$-dependence of which thus becomes significant for $h \sim {\cal O}(\frac{U}{2})$ (albeit that satellite shifts are naturally perceptible in figure~2 for $h \sim {\cal O}(\delno)$).
The position of the Hubbard satellites is however secondary: the $h$-dependence of their widths is the significant issue, as now explained.

Upon addition of a $\da$-spin electron to an $\ua$-spin occupied impurity, with energy cost $\sim \frac{U}{2} + h$ corresponding to the position of the upper Hubbard satellite, two subsequent hopping processes may occur.  The added $\da$-spin may hop off the impurity, itself leading to an electron loss rate (and hence HWHM spectral broadening) of $\delno$.  This is elastic scattering, and is all that is captured at the one-electron MF level.  Alternatively the $\ua$-spin electron already present may hop off the impurity, leading to a spin-flip with energy cost of order $\wmh$; and hence to an additional loss rate of $\delno$ provided $\wmh \ll \frac{U}{2} + h$.  This is a many-body process that has no counterpart at MF level.
The total loss rate of electrons from the site is thus $2\delno$, whence the HWHM of the Hubbard satellites is doubled (and their peak intensity correspondingly halved) compared to the simple MF result.
This behaviour is clearly evident in figure~2 (where $\pdod \ofwh \simeq \frac{1}{4}$ at the peak positions of the Hubbard satellites); its formal origins within the LMA for $h=0$ have been discussed in reference~\cite{ref:L1}.

For sufficiently large fields however, $\wmh$ will become comparable to $\frac{U}{2} +h$ and the spin-flip energy cost effectively prohibits the additional many-body broadening.  Since $\wmh \sim 2h$ for $h \gg \wmo$, this arises for $h \sim \frac{U}{2}$.  That the additional many-body broadening is `switched off' for fields of this order is indeed seen in figure~2$b$ for $h / \wmo = 1.2 \times 10^4$ and $1.6 \times 10^4$ ($h / \left[\frac{U}{2}\right] \simeq 1.5$ and $2$ respectively); and $D \ofwh$ is given asymptotically by

\be
\label{eq:bighd}
D \ofwh = \frac{1}{2} \sum\limits_\sigma \frac{\delno \pi^{-1}}{\left[ \w + \sigma\left( \frac{U}{2} + h \right)\right]^2 + \delno^2}
\ee

\noindent such that $\pdod \ofwh \simeq \frac{1}{2}$ for $|\w| = \frac{U}{2} + h$.

Equation~(\ref{eq:bighd}) is of course the pure MF spectrum in strong coupling, and that it is asymptotically exact for $h \gg \frac{U}{2}$ may be seen from simple consideration of the $h$-dependent one-electron site energies (see equation~(\ref{eq:ham})), viz $\epsilon_{i\sigma} = \epsilon_i - \sigma h$ with $\epsilon_i = -\frac{U}{2}$ for the symmetric AIM.  If $\epsilon_{i\da} \gg 0$ and $\epsilon_{i\ua} \ll 0$ (with $0$ the Fermi level) -- \ie if $h \gg |\epsilon_i | = \frac{U}{2}$ -- then in the ground state the impurity is occupied only by $\ua$-spin electrons.  As far as $D_\ua \ofwh$ is concerned only $\ua$-spin electrons are then involved in virtual hopping processes; so the fermions are effectively spinless, leaving an effective one-body problem with corresponding site energy $\epsilon_i^{\rm eff} = \epsilon_i - h$.  $D_\ua \ofwh$ is then a Lorentzian of width $\delno$  centred on $\w = \epsilon_i^{\rm eff}$($= -[\frac{U}{2} +h]$), and $D_\da \ofwh$ follows from p-h symmetry.
Equation~(\ref{eq:bighd}) results for $D\ofwh$.  It is the spectral signature of the free-orbital regime where the static impurity susceptibility $\chi_\rmi (h)$ coincides with that of the $U = 0$ limit (as is captured by the LMA, see \cite{ref:ld}).  The crossover to such behaviour should be observable in NRG calculations of $D \ofwh$, for although NRG cannot handle adequately the many-body broadening of the Hubbard satellites for $h \gg \frac{U}{2}$,  one-electron broadening arising from the hybridization $\Delta(\w)$ is well captured by a technique introduced recently by Bulla, Hewson and Pruschke \cite{ref:bulla2}.

\subsection{Approach to the Kondo scaling limit.}

In strong coupling $\ut \gg 1$, the low-energy physics of the AIM depends solely upon the Kondo scale.  The latter itself appears in a variety of superficially different guises, viz $\wmo$, $\wk$ or $\delno Z(0)$ (with $Z(0)$ the zero-field quasiparticle weight, see equation~(\ref{eq:zhdef})).  But these are all of course equivalent, being simply proportional to each other (\eg $\delno Z (0) = \frac{\pi}{4} \wmo$ and $\wk = 0.691 \wmo$ within the present LMA \cite{ref:dl}); we denote any of them by $\wa$.

Although $\wa$ itself depends upon the interaction strength ($\wa \propto \exp (-\pi U / 8 \delno)$), the fact that it is the sole low-energy scale in SC means that the Kondo/Abrikosov-Suhl resonance exhibits universal scaling in terms of $\w / \wa$ alone, with no explicit dependence on the bare material parameters.  That the LMA for $h=0$ leads to such scaling behaviour with progressively increasing $\ut$ has been shown in reference~\cite{ref:L1}; the resultant scaling spectrum in the Kondo limit has also been obtained analytically in reference~\cite{ref:dl}, and shown to give good agreement with $h=0$ NRG calculations \cite{ref:dl,ref:bulla1}.

Scaling behaviour should likewise arise for $h \ne 0$, but with the scaling spectrum $\pdod \ofwh \equiv F(\w / \wa ; h / \wa)$ now dependent upon $\w / \wa$ and $h / \wa$.  That such behaviour arises within the LMA upon progressively increasing $\ut$ is illustrated in figure~3.  We consider a fixed value of $\wmh / \wmo = 21$, corresponding from equation~(\ref{eq:scwmh}) to a fixed $\hp = h / \wmo = 10$ in the Kondo limit where $|\mu_0| \ra 1$ (and to  $\hp$ within 10\% of this value for $\ut \gtrsim 4$ - $5$).  For three different interaction strengths $\ut = 2$, $4$ and $8$, the resultant $\pdod \ofwh$ are shown versus $\w / \wmh$ ($\propto \w / \wmo$).  They are indeed seen to approach asymptotically the Kondo scaling limit (albeit somewhat more slowly than for $h=0$ \cite{ref:L1}).  The latter is also shown in figure~3, and will now be investigated in detail.

\begin{figure}
\centering\epsfig{file=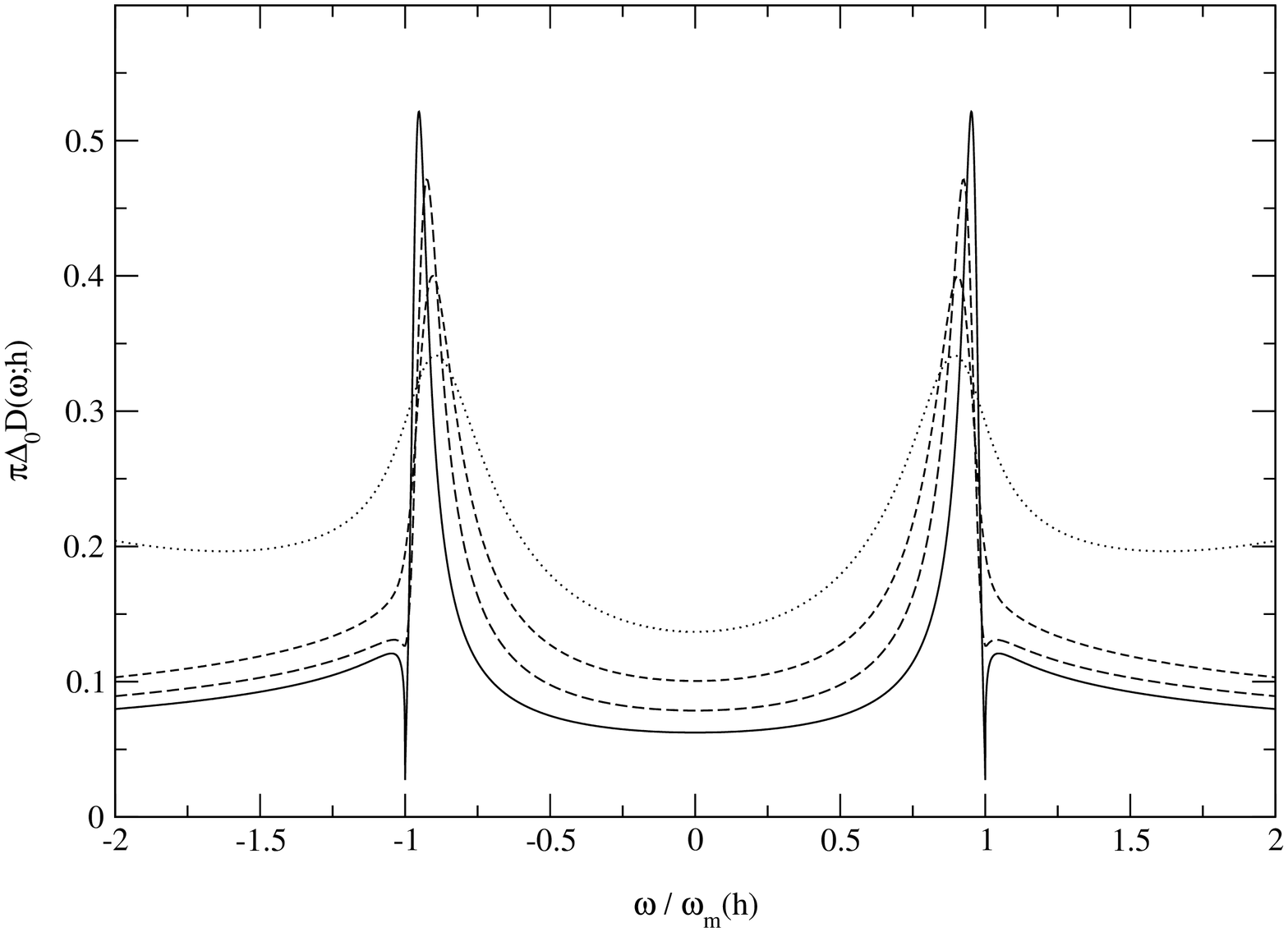,width=100mm,angle=270}
\vskip-5mm
\protect\caption{$\pdod \ofwh$ vs $\wt = \w / \wmh$ for fixed $\wmh / \wmo = 21$ (\ie $h / \wmo \simeq 10$), and $\ut = 2$ (dotted line), $4$ (short dash) and $8$ (long dash).  The Kondo limit scaling spectrum obtained in \S 5.1 is also shown (solid line).}
\label{fig:3}
\end{figure}

\section{Scaling spectrum: general considerations.}
\label{sec:4}

Our aim here is to obtain analytically the $h$-dependent LMA scaling spectrum appropriate to the Kondo limit; and in particular to do so with only minimal assumptions about the form of the transverse spin polarization propagator $\ppm \ofwh$ that enters the dynamical part of the self-energy (equation~(\ref{eq:sigint})).  As for the zero-field problem considered in reference~\cite{ref:dl} the latter will enable us to encompass, and go beyond, the particular case considered in the previous section where $\ppm \ofwh$ is given by the p-h ladder sum.

To obtain the scaling behaviour for the Kondo resonance, one considers finite $\w / \wmo$ and finite $\hp = h / \wmo$ in the limit $\wmo \propto \exp (-\pi U / 8 \delno) \ra 0$; the latter projects out the non-universal features that are not part of the scaling spectrum (such as the Hubbard satellites).  Hence, referring to equations~(\ref{eq:g_to_gsig}, \ref{eq:gsig}), the `bare' $\w = [\w / \wmo]\wmo \equiv 0$ may be neglected, as too may the bare field $h = \hp \wmo \equiv 0$; and $\Delta(\w)$ likewise reduces to $\Delta(0) = -\rmi~\sgn(\w)\delno$.  The scaling spectrum then follows from equations~(\ref{eq:g_to_gsig}, \ref{eq:gsig}) as

%how to move left margin within IOPtex eqn environment?
\be
\label{eq:scdh}
\hspace{-2.5cm}\pdod \ofwh = \frac{1}{2} \sum\limits_\sigma \! \frac{\left[ 1  + \frac{1}{\delno} \sigsi \ofwh \right]^2}{\left[ \frac{1}{\delno}\!  \left(\sigtsr \ofoh \! +\!  \left[ \sigsr \ofwh\!  -\!  \sigsr\ofoh \right] \right) \right]^2\!  +\!  \left[ 1\!  +\!  \frac{1}{\delno} \sigsi \ofwh \right]^2      }
\ee

\noindent (where $\sigtsi \ofwh = \sigsi \ofwh$ and $\sigtsr \ofwh - \sigtsr\ofoh = \sigsr \ofwh - \sigsr\ofoh$ are trivially used).  Equation~(\ref{eq:scdh}) is general, in the sense that provided the host is metallic it applies to any one-electron hybridization $\Delta(\w)$.

As expected, the scaling behaviour of the spectrum is thus determined exclusively by that of the self-energies.  And the form of the latter is in turn closely related to the corresponding zero-field problem considered in reference~\cite{ref:dl}.  Specifically, the transverse spin polarization propagator that enters the LMA $\sigs \ofwh$ (equation~(\ref{eq:sigint})) has the following functional form in SC,

\be
\label{eq:scpiform}
\frac{1}{\pi} \im\ppm \ofwh = \frac{A}{\wmh} f(\wt) \theta(\wt)
\ee

\noindent with

\be
\label{eq:pinorm}
\int\limits^\infty_0 \frac{\rmd\w}{\pi}~ \im \ppm \ofwh = 1 = A \int\limits^\infty_0 \rmd y~f(y)
\ee

\noindent and $\wt = \w / \wmh$.  The essential point here is that $\im\ppm\ofwh$ for $h > 0$ has the same functional form as for $h=0$ \cite{ref:dl}, scaling in terms of $\wt = \w / \wmh$ in the same way as it does in terms of $\w / \wmo$ for $h=0$.  Such behaviour is physically natural, since $\wmh$ remains the sole low-energy spin-flip scale for $h \ne 0$, just as it is for $h = 0$ where $\wmo = \wm(h=0)$.  Three further points should be noted here.  First, equation~(\ref{eq:pinorm}) embodies physically the saturation of the local moment in SC ($|\mu| \ra 1)$; the constant A being $h$-independent and determined by the resultant `normalization'.  Second, by definition of $\wmh$, the function $f(\wt)$ is peaked at $\wt =1$; and $f(\wt) \sim \wt$ as $\wt \ra 0$.  Finally, we add that the above behaviour is readily shown to arise explicitly with $\ppm \ofwh$ given by the p-h ladder sum; from which $f(\wt)$ is found to have the form \cite{ref:L1,ref:dl}

\be
\label{eq:lmafw}
f(\wt) = \frac{\wt}{1 - 2\alpha \wt + \wt^2}.
\ee

\noindent In the following however, the particular functional form of $f(\wt)$ will not be required; we shall need it only in \S 5.  

Using equation~(\ref{eq:scpiform}), the scaling behaviour of $\sigs \ofwh$ (equation~(\ref{eq:sigint})) follows in direct parallel to the $h=0$ case \cite{ref:dl}.  We thus quote only the relevant results from reference~\cite{ref:dl}.  Specifically

\be 
\label{eq:scsigih}
\delno^{-1} \sigui \ofwh = \theta (-\wt) 4 A \int\limits^{|\wt|}_{0} \rmd y ~  f(y)
\ee

\noindent (and $\sigdi \ofwh = \sigui (-\w;h)$), which is required in equation~(\ref{eq:scdh}) and scales solely in terms of $\wt = \w / \wmh$ with no explicit $h$- or $\ut$-dependence.  And for $\sigur \ofwh$ ($= -\sigdr (-\w;h)$),

\be
\label{eq:scsigrh}
\delno^{-1} \sigur \ofwh = \frac{4}{\pi} \ln \left[ \frac{\lambda}{\wmh} \right] - \frac{4}{\pi} A \int\limits^\infty_0 \rmd y~f(y)\ln|y + \wt|
\ee

\noindent where (see \S 2) $\lambda = {\rm min}[D,\frac{U}{2}]$.  From this the quasiparticle weight, given generally by equation~(\ref{eq:zhdef})and reducing to $Z(h)^{-1} = -(\partial \sigsr \ofwh / \partial \w)_{\w = 0}$ in the SC scaling regime, is thus given by

\be
\label{eq:sczhdef}
\frac{1}{\delno Z(h)} = \frac{1}{\wmh} \frac{4A}{\pi} \int\limits^{\infty}_0 \rmd y~\frac{f(y)}{y}
\ee

\noindent and satisfies

\be
\label{eq:hratios}
\frac{\zh}{\zo} = \frac{\wmh}{\wmo}
\ee

\noindent (since $A$ is $h$-independent).  Equation~(\ref{eq:scsigrh}) also yields directly

\be
\label{eq:scsigrdiff}
\delno^{-1} (\sigur \ofwh - \sigur \ofoh) = -\frac{4}{\pi} A \int\limits^\infty_0 \rmd y~f(y)\ln \left| 1 + \frac{\wt}{y}\right|
\ee

\noindent which is likewise required in equation~(\ref{eq:scdh}) for $\pdod \ofwh$; and which again scales solely in terms of $\wt$.

The explicit $U$-dependence of the zero-field Kondo scale $\wmo \equiv \wm (h=0)$ follows from symmetry restoration equation~(\ref{eq:sigpin}), viz $\sigur (0;0) = \frac{1}{2} U$ (since $|\mu (0) | \ra 1$ in SC).  Equation~(\ref{eq:scsigrh}) for $h=0$ then yields directly

\numparts
\be
\label{eq:wmp2wm0}
\wmo = c \wmp
\ee

\noindent with $c$ a $U$- and $h$-independent constant of order unity given by

\be
\label{eq:cdef}
c = \exp \left[ -A \int\limits^\infty_0 \rmd y ~ f(y) \ln (y)\right]
\ee

\noindent and

\be
\label{eq:wmpdef}
\wmp = \lambda \exp \left[ \frac{-\pi U}{8\delno}\right].
\ee
\endnumparts

\noindent The remaining quantity required to determine equation~(\ref{eq:scdh}) for $\pdod \ofwh$ is $\delno^{-1}\sigtsr \ofoh$ (equation~(\ref{eq:sig_to_sigt})); which reduces in SC to $\sigtur \ofoh = -\frac{U}{2} + \sigur \ofoh = \sigur \ofoh - \sigur (0;0)$, again using $|\mu(0)| \ra 1$ together with $\delta |\mu (h)| = 0$ (as follows from equation~(\ref{eq:scdmu}) since the bare $h = \hp\wmo \equiv 0$).  Since $A$ is $h$-independent, equation~(\ref{eq:scsigrh}) then yields $\delno^{-1} \sigtur \ofoh = -\frac{4}{\pi}\ln (\wmh / \wmo)$; \ie using equation~(\ref{eq:hratios}),

\be
\label{eq:scsigr0}
\delno^{-1} \sigtur \ofoh = - \frac{4}{\pi} \ln \left[ \frac{Z(h)}{Z(0)} \right]
\ee

\noindent (with $\sigtdr \ofoh = - \sigtur \ofoh$).

Our final task is to determine the explicit field-dependence of  $\zh /\zo$.  This may be deduced in two distinct ways.  First,  and more generally, using an exact Ward identity pertaining to the Kondo regime of vanishing charge fluctuations \cite{ref:hewson}:

\be
\label{eq:ward}
\frac{\partial \sigtur \ofoh}{\partial h} = \frac{-2}{\zh} + 1.
\ee

\noindent Combined with equation~(\ref{eq:scsigr0}), this leads to a trivial differential equation for $\zh$; solution of which yields the LMA result for $\zh / \zo$, viz

\be
\label{eq:zhz0}
\frac{\zh}{\zo} = 1 + \frac{\pi}{2}\frac{h}{\delno \zo}
\ee

\noindent (remembering that we consider finite $h / \delno \zo$ and $\delno\zo \propto \wmo \ra 0$).  Alternatively, the $h$-dependence may be deduced directly from the LMA with $\ppm \ofwh$ given explicitly by the p-h ladder sum; which we refer to from now on as the LMA(RPA).  In this case, as discussed in \S 3, the spin-flip scale $\wmh / \wmo = 1 + 2 h / \wmo$ in SC.  But for the LMA(RPA), $Af(y) = \delta(y-1)$ (see \cite{ref:dl} and \S 5 below), and thus $\delno \zo  = \frac{\pi}{4}\wmo$ from equation~(\ref{eq:sczhdef}); hence using equation~(\ref{eq:hratios}), equation~(\ref{eq:zhz0}) for $\zh / \zo$ is again recovered.  Equation~(\ref{eq:zhz0}) is of course approximate.  An exact result for $\zh / \zo$ can however be deduced using the Ward identity equation~(\ref{eq:ward}); and comparison of which to equation~(\ref{eq:zhz0}) shows the latter to be a good approximation over essentially the entire range of field strengths (figure~\ref{fig:zcomp}).

For any finite field $h / \delno \zo$, the scaling behaviour of the single-particle spectrum $\pdod \ofwh$ (equation~(\ref{eq:scdh})) may now be obtained directly using equations~(\ref{eq:scsigih}, \ref{eq:scsigrdiff}, \ref{eq:scsigr0}, \ref{eq:zhz0}) together with p-h symmetry.  This will be considered explicitly in \S 5 for a particular form of the function $f(\wt)$ that determines (see equation~(\ref{eq:scpiform})) the transverse spin polarization propagator.  First however,  we consider predictions arising from the preceding analysis that are essentially independent of the details of $f(\wt)$.

\subsection{Spectral Limits}

We here consider briefly the field-dependence of the single-particle spectrum at the Fermi level $\w = 0$, as well as the behaviour of the spectral tails for frequencies $|\wt| = |\w| / \wmh \gg 1$.

The Fermi level spectrum is given from equations~(\ref{eq:g_to_gsig}, \ref{eq:gsig}) by $\pdod \ofoh = [ 1 + ([h - \sigma \sigtsr \ofoh ]/\delno)^2 ]^{-1}$ (with $\sigma \sigtsr \ofoh$ independent of $\sigma$ by p-h symmetry).  
Defining $\htilde = h / \delno \zo$, the LMA result for $\pdod \ofoh$ in the Kondo limit of finite $\htilde$ and $\delno \zo \ra 0$ thus follow from equations~(\ref{eq:scsigr0}, \ref{eq:zhz0}) as

\be
\label{eq:scd0}
\pdod \ofoh = \left[ 1 + \left( \frac{4}{\pi} \ln \left[ 1 + \frac{\pi}{2}\htilde\right] \right)^2 \right] ^{-1},
\ee

\noindent with asymptotic field dependencies:

\numparts
\be
\label{eq:scd0lh}
\pdod \ofoh ~~ {_{\htilde \ll 1}\atop^{\sim}}~~  1 - 4 \htilde^2
\ee

\be
\label{eq:scd0sh}
\pdod \ofoh ~~ {_{\htilde \gg 1}\atop^{\sim}}~~  \left[ \frac{4}{\pi} \ln (\htilde)\right]^{-2}.
\ee

\endnumparts

\noindent The exact behaviour of $\pdod \ofoh$ can be obtained, since the (excess) impurity magnetization $M_{\rmi} (h)$ follows generally from the Friedel sum rule as \cite{ref:hewson} $\pi M_\rmi (h) / \gmb = \tan^{-1}[(h - \sigma \sigtsr \ofoh ) / \delno] $; whence

\be
\label{eq:scmag}
\pdod \ofoh = \cos ^2 \left[ \frac{\pi M_\rmi (h)}{\gmb}\right]
\ee

% hard ref
\noindent with $M_\rmi (h)$ known from the Bethe Ansatz (BA) solution of the Kondo/s-d model \cite{ref:afl}.  From this the exact low- and high-field asymptotics of $\pdod \ofoh$ are readily obtained, and are given by equations~(4.15).  The LMA thus captures these correctly, which is non-trivial.

\begin{figure}
\centering\epsfig{file=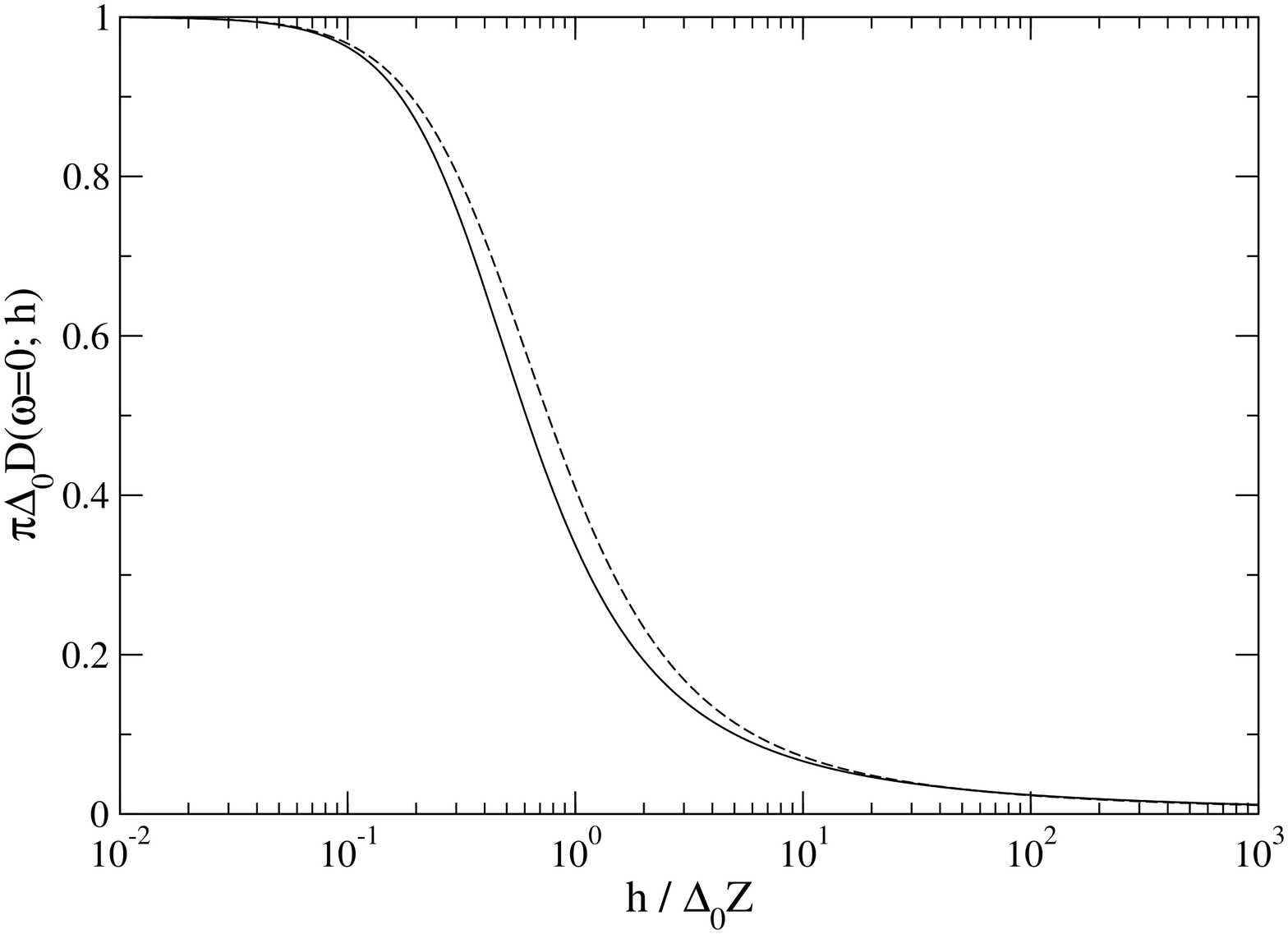,width=100mm,angle=270}
\vskip-5mm
\protect\caption{$\pdod (\w = 0;h )$ vs $\htilde = h / \delno \zo$ obtained from the LMA (dashed line, equation~(\ref{eq:scd0})), compared to the exact result from the Bethe ansatz \cite{ref:afl} (solid line).}
\label{fig:4}
\end{figure}

A full comparison between the LMA and exact BA results for $\pdod \ofoh$ is given in figure~\ref{fig:4}, and the level of agreement for all field strengths is self-evident.  We also add that numerical renormalization group calculations of the single-particle spectrum for the Kondo model \cite{ref:costi} yield excellent agreement with the BA result for $\pdod \ofoh$, over the modest range of field strengths considered in reference~\cite{ref:costi}.  Static properties arising from the LMA are discussed further in reference~\cite{ref:ld}.  In particular the Wilson ratio $\rw (h) = 2~ \forall~ h$ \cite{ref:tw,ref:wf} is correctly recovered; and the field dependences of $M_\rmi (h)$ and the corresponding static susceptibility $\chi_\rmi (h)$ are likewise found to be asymptotically exact in both the weak and strong field limits.

The behaviour of the spectrum for frequencies $|\wt| = |\w| / \wmh \gg 1$ is also readily obtained.  From equations~(\ref{eq:scsigih}, \ref{eq:scpiform}), $\delno^{-1} \sigui \ofwh$ is given asymptotically by $\delno^{-1} \sigui \ofwh = 4 \theta (-\wt)$.  Equation~(\ref{eq:scsigrdiff}) for $|\wt| \gg 1$ yields $\delno^{-1} ( \sigur \ofwh - \sigur \ofoh ) = -\frac{4 A}{\pi} \int\nolimits^\infty_0 \rmd y ~ f(y) \ln [|\wt| / y] = -\frac{4}{\pi} \ln [ |\wt| c]$ where equation~(\ref{eq:pinorm}) is used and the constant $c$ is given by equation~(\ref{eq:cdef}); and $\delno ^{-1} \sigtur \ofoh = -\frac{4}{\pi} \ln [\wmh / \wmo]$ from equations~(\ref{eq:scsigr0}, \ref{eq:hratios}).  Hence $\sigtur \ofwh = \sigtur \ofoh + [\sigur \ofwh - \sigur \ofoh]$ is given for $|\wt| \gg 1$ by $\delno ^{-1} \sigtur \ofwh = -\frac{4}{\pi} \ln (|\w|/\wmp)$ where equation~(\ref{eq:wmp2wm0}) is used, and the $\ut$-dependence of $\wmp$ is given explicitly by equation~(\ref{eq:wmpdef}).  The scaling spectrum for $|\wt| \gg 1$ then follows from equation~(\ref{eq:scdh}) as

\be
\label{eq:sclargewd}
\pdod \ofwh ~~ {_{|\wt| \gg 1}\atop^{\sim}}~~ \frac{1}{2} \left\{ \frac{1}{\left[\frac{4}{\pi}\ln |\wp|\right]^2 +1} + \frac{5}{\left[\frac{4}{\pi}\ln |\wp|\right]^2 +25} \right\}
\ee

\noindent where $\wp = \w / \wmp$; equivalently, the frequency dependence may be recast in terms of $\w / \wk$ (with $\wk$ the HWHM of $D(\w;0)$), using $\wk / \wmp = 0.691$ found for the $h=0$ LMA \cite{ref:dl}.

Equation~(\ref{eq:sclargewd}) is precisely the result for the behaviour of the scaling spectrum `tails' obtained in reference~\cite{ref:dl} for $h=0$, and there formally applicable for $|\w| \gg \wmo = \wm (h=0)$.  The resultant slowly varying tail is logarithmic in form, as opposed to the algebraic Doniach-${\rm \breve{S}unji\acute{c}}$ \cite{ref:ds} decay $D(\w;0) \sim (|\w| / \wk)^{-\frac{1}{2}}$ hitherto thought to arise (see \eg \cite{ref:frota,ref:silver,ref:bulla1}).  We have however argued \cite{ref:dl} that the logarithmic decay is entirely natural in physical terms, and shown both that equation~(\ref{eq:sclargewd}) gives excellent agreement with NRG calculations of $D(\w;0)$ (see \eg figure~2 of reference~\cite{ref:dl}); and that this long tail in fact dominates the $h=0$ scaling spectrum, the crossover to Fermi liquid form occurring only on the lowest energy scales $|\w|/\wk \ll 1$.  The arguments leading to equation~(\ref{eq:sclargewd}) show additionally that this tail behaviour is asymptotically common both to $h=0$ and $h > 0$, where it arises for $|\wt| = |\w| / \wmh \gg 1$; \ie for $|\w| / \wmo \gg 1 + \frac{\pi}{2}\htilde$ (using equations~(\ref{eq:hratios}, \ref{eq:scsigr0})).  This will be seen explicitly in the results shown in the following section.

\section{LMA Scaling spectrum.}
\label{sec:5}

To obtain the LMA scaling spectrum on all energy scales requires a full specification of the transverse spin polarization propagator (equation~\ref{eq:scpiform}).  This is now considered, first (\S 5.1) using the LMA(RPA) with $\ppm \ofwh$ given explicitly by the p-h ladder sum; and then (\S 5.2) via a simple modification thereof introduced previously in reference~\cite{ref:dl}.  In \S 5.3 we consider the spinon approximation to the $h$-dependent single-particle spectra developed in reference~\cite{ref:mw}.  Comparison to recent transport experiments on quantum dots in the Kondo regime \cite{ref:qdots} is made in \S 5.4.

\subsection{LMA(RPA).}

For the LMA(RPA) the function $f(\wt)$ that determines $\im \ppm \ofwh$ (equation~(\ref{eq:scpiform})) has the form equation~(\ref{eq:lmafw}), where the $x$-dependence of $\alpha$ (and $A$, see equation~(\ref{eq:pinorm})) is given explicitly  in reference~\cite{ref:L1}.  From this it is known \cite{ref:dl} that in SC $\ut \gg 1$, $\alpha \ra 1$ and $A \sim [2(1-\alpha)]^{\frac{1}{2}}/\pi \ra 0$ such that $A f(y) = \delta(y-1)$ \ie from equation~(\ref{eq:scpiform}) $\frac{1}{\pi} \im \ppm \ofwh = \delta (\w - \wmh)$ reduces to a delta function centred on $\wmh = \wmo + 2h$.  Note also that $\wmo = \wmp$ follows from equations~(\ref{eq:wmp2wm0}, \ref{eq:cdef}), with the $\ut$-dependence of $\wmp$ given explicitly by equation~(\ref{eq:wmpdef}).

From equations~(\ref{eq:scsigih}, \ref{eq:scsigrdiff}) it follows directly that

\numparts
\bea
\label{eq:scrpasigr}
\delno^{-1} ( \sigur \ofwh - \sigur \ofoh) &= -\frac{4}{\pi} \ln | \wt +1|\\
\label{eq:scrpasigi}
\delno^{-1} \sigui \ofwh &= 4 \theta (-[\wt +1])
\eea

\endnumparts

\noindent where $\wt = \w / \wmh $; while $\delno^{-1} \sigtur \ofoh = -\frac{4}{\pi} \ln [ \wmh / \wmo]$ from equations~(\ref{eq:scsigr0}, \ref{eq:hratios}).  The spectrum in closed form then follows simply from equation~(\ref{eq:scdh}); specifically for $\w > 0$ (since $D(-\w;h) = D \ofwh$) by:

\bea
\label{eq:scrpad}
\nonumber\hspace{-2.5cm}\pdod \ofwh = \\
\vspace{0.4cm}
\hspace{-1.75cm}\frac{1}{2} \left\{ \frac{1}{\left[ \frac{4}{\pi} \ln | \wp + \wmph| \right]^2 +1} + \frac{1 + 4\theta ( \wp - \wmph)}{\left[ \frac{4}{\pi}\ln | \wp - \wmph| \right]^2 + \left[ 1 + 4 \theta (\wp - \wmph) \right]^2}\right\}
\eea

\noindent Here $\wp = \w / \wmp$, and likewise $\wmph  = \wmh / \wmp = 1 + 2 \hp$ where $\hp = h / \wmp$ ($\equiv h / \wmo $ as used in \S 3); equivalently $\wmph = 1 + \frac{\pi}{2} \htilde$, where (as in \S 4) $\htilde = h / \delno \zo$ and $\delno \zo = \frac{\pi}{4} \wmp$ for the LMA(RPA).  Equation~(\ref{eq:scrpad}) is the Kondo limit scaling spectrum shown in figure~\ref{fig:3} above (for fixed $\wmph = 21$) and arising as the large-$\ut$ limit of the AIM.  It naturally reduces in the zero-field limit (where $\wmp (0) = 1$) to the result obtained  previously in reference~\cite{ref:dl}.

\begin{figure}
\centering\epsfig{file=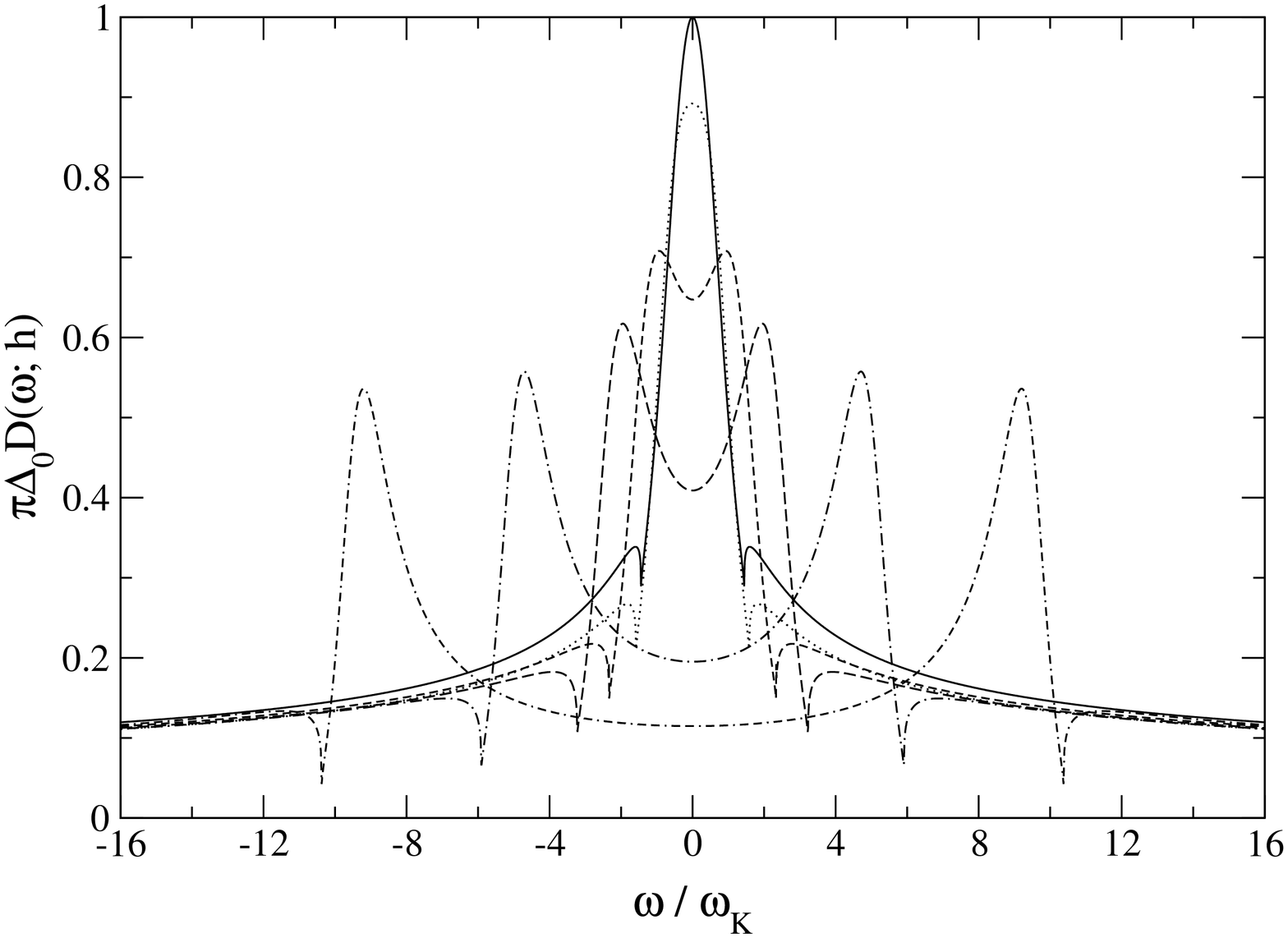,width=100mm,angle=270}
\vskip-5mm
\protect\caption{LMA(RPA) scaling spectrum $\pdod \ofwh$ vs $\w / \wk$ for $\htilde = h / \delno \zo = 0$, $0.2$, $0.5$, $1$, $2.5$ and $5$ (from top to bottom).}
\label{fig:5}
\end{figure}

The field dependence of the resultant spectrum is shown in figure~\ref{fig:5}: $\pdod \ofwh$ vs $\w / \wk$  for $\htilde = h / \delno \zo$ = 0, 0.2, 0.5, 1, 2.5 and 5; here as throughout, $\wk$ is defined as the HWHM of the zero-field spectrum (with $\wk / \wmp = 0.691$ for the LMA \cite{ref:dl}).  The spectrum at the Fermi level, $\w =0$, decreases monotonically with increasing field (figure~\ref{fig:4}); and for sufficiently small fields $(\partial ^2 D \ofwh / \partial \w ^2)_{\w=0} < 0$, whence the maximum in the Kondo resonance remains at $\w = 0$.  As pointed out by Costi in a recent NRG study  of the Kondo model \cite{ref:costi} however, the Kondo resonance `splits' above a certain field $H_{\rm c}$  where the $\w = 0$ curvature changes sign.  For the LMA(RPA) this may be obtained analytically, and occurs at $2 h_{\rm c} / \wk = \gmb H_{\rm c} / \wk = 0.600$; compared to the corresponding NRG value \cite{ref:costi} of $0.5$.

With increasing $\htilde > \htilde_{\rm c}$ the split Kondo peaks in $D \ofwh$ move further apart, although as seen (figure~\ref{fig:5}) the generic tail behaviour equation~(\ref{eq:sclargewd}) is always approached at sufficiently high frequencies.  The LMA(RPA) peak splitting is readily deduced from the individual $D_\sigma \ofwh$ (with $\pdod _\da \ofwh$ given in its entirety by the second term in equation (\ref{eq:scrpad})): since $\wmph = 1 + 2 \hp$, it follows directly using equation~(\ref{eq:scrpad}) that $D_\sigma \ofwh = D_\sigma(\w + 2\sigma h; 0)$ with $D_\sigma (\w;0)$ the zero-field case.  $D_\sigma \ofwh$ thus amounts simply to a rigid shift of its zero-field counterpart, and hence has its peak maximum at $\w_{\rm p}^\prime = - 2 \sigma \hp$ precisely.  The corresponding maxima in $D \ofwh$ ($= \frac{1}{2} \sum\nolimits_\sigma D_\sigma \ofwh$) for $\hp > \hp_{\rm c}$ likewise occur in practice very close to $|\w_{\rm p}| = 2 \hp$, figure~\ref{fig:5} (and asymptotically so for $\hp \gg 1$); with a total peak splitting of $4h$, or twice the Zeeman energy.  We add that a peak position of $|\w_{\rm p}| = 2h$ in $D_\sigma \ofwh$ arises as the large-field ($h \gg \wk$) asymptotic behaviour within the spinon approximation \cite{ref:mw} (discussed further in \S 5.3).  It also arises within an equation of motion approach \cite{ref:mwl}; although the apparent parallel here is somewhat misleading since the Kondo scale does not exist within such an approach, and hence neither does the Kondo scaling regime of finite $h / \wk$ and $\w / \wk$.

\subsection{Beyond the LMA(RPA).}

While the LMA(RPA) captures well the spectrum at the Fermi level (figure~\ref{fig:4}), and probably also the spectral tails (as for $h=0$ \cite{ref:dl}, see also figure~\ref{fig:walter} below), its quantitative limitations are apparent from figure~\ref{fig:5}.  The split Kondo peaks, for example, are too pronounced in comparison to recent NRG calculations for the Kondo model \cite{ref:costi}.  The origins of this limitation stem from the divergence in the LMA(RPA) $\sigsr \ofwh$ at $|\wt| =1$ (equation~(\ref{eq:scrpasigr})), reflected by the `dip' in $\pdod \ofwh$ at $\w = \wmh$ evident in figure~\ref{fig:5}.  As known for the $h=0$ problem \cite{ref:L1,ref:dl}, this is entirely an artifact of the specific RPA-like form for the polarization propagator, embodied in the fact that the resultant $\im \ppm \ofwh$ is a $\delta$-function at $\w = \wmh$: in reality one expects $\im \ppm \ofwh$ to have a finite width.

To rectify this deficiency we proceed as in reference~\cite{ref:dl}.  We retain the form equation~(\ref{eq:lmafw}) for $f(\wt)$, which has a finite width provided $\alpha \ne 1$; and we employ a high-frequency cutoff $\wct$ to render $f(\wt)$ normalizable (equation~(\ref{eq:pinorm})).  The width parameter $\alpha$ is then determined by requiring that the leading low-frequency behaviour ($\propto \w^2$) of the imaginary part of the conventional single self-energy for $h=0$ is recovered exactly.  As discussed in reference~\cite{ref:dl} this requires $A = \frac{1}{2} [ \wmo / \delno \zo ]^2$ ($=\frac{1}{2} [ \wmh / \delno \zh ]^2$, see equation~(\ref{eq:hratios})), which via equations~(\ref{eq:pinorm}, \ref{eq:sczhdef}) in turn implies a simple equation that determines $\alpha$ uniquely  for the chosen cutoff $\wct$.  The latter is of course arbitary but, as expected physically, results are not sensitive to it \cite{ref:dl}; in practice, as in reference~\cite{ref:dl}, we choose $\wct = 10$ (and hence $\alpha = 0.308$).

\begin{figure}
\centering\epsfig{file=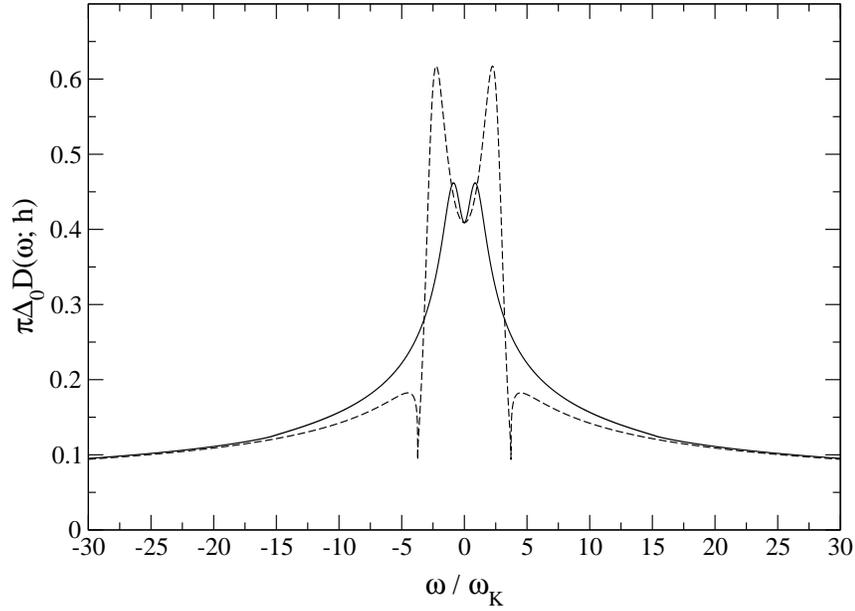,width=100mm,angle=270}
\vskip-5mm
\protect\caption{For $\htilde = 1$, comparison between scaling spectra $\pdod \ofwh$ arising from the LMA(RPA) (dashed line) and the LMA (solid line) discussed in text.}
\label{fig:6}
\end{figure}

While the effects of this modification are rather minor in comparison to the LMA(RPA) for $h=0$ \cite{ref:dl}, they are more significant at finite fields.  For $\htilde = h / \delno \zo = 1$, figure~\ref{fig:6} shows $\pdod \ofwh$ vs $\w / \wk$ for the LMA(RPA) compared to the LMA outlined above (and refered to simply as the LMA).  The zero-frequency and asymptotic tail behaviours of the two naturally coincide, being independent of the detailed form of $f(\wt)$ (see \S 4.1).  But elsewhere the differences are clearly quite significant.  In particular the Kondo peaks in the LMA are less pronounced, and for the example shown in figure~\ref{fig:6} their splitting is smaller in comparison to the LMA(RPA).

\begin{figure}
\centering\epsfig{file=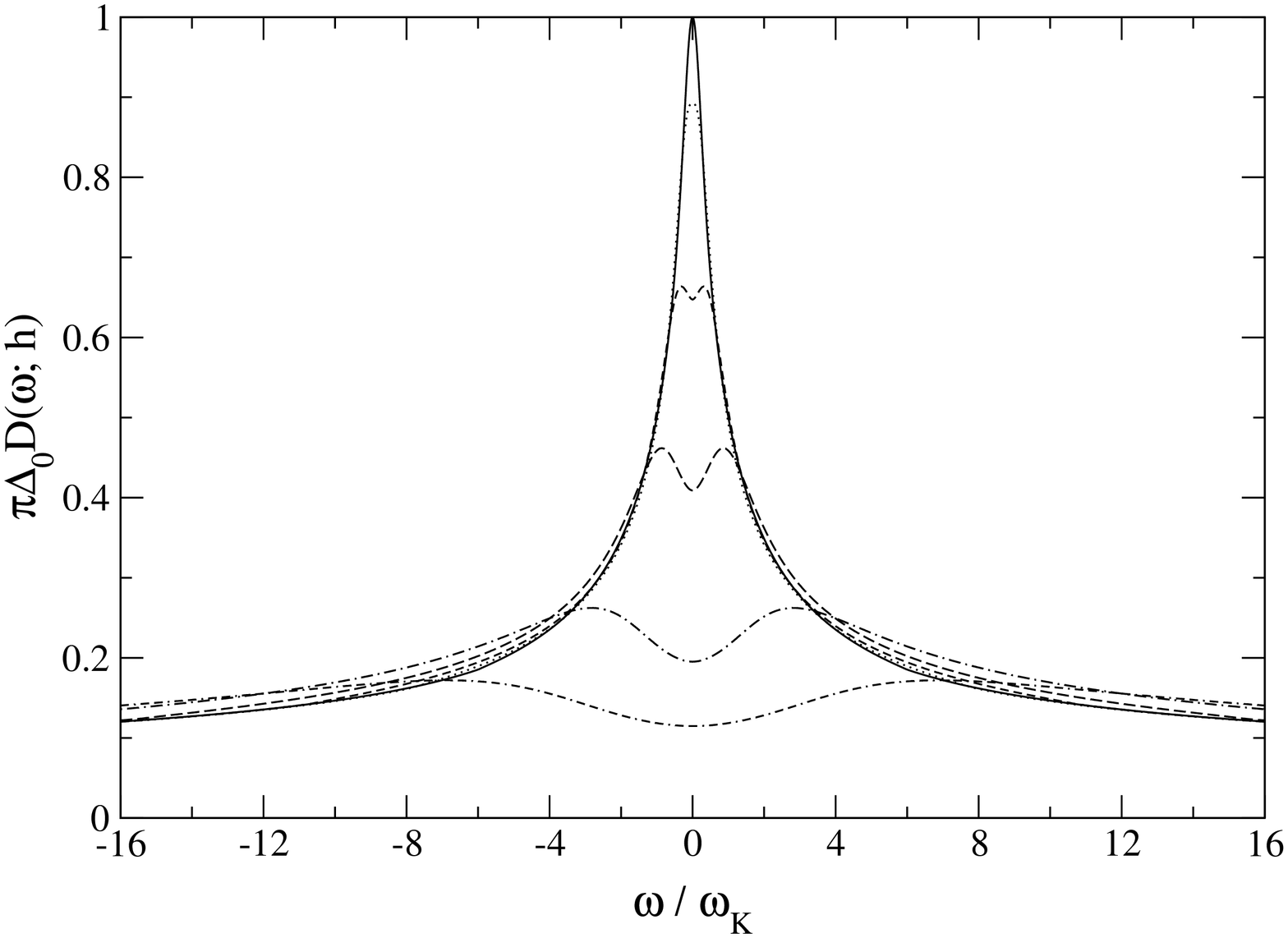,width=100mm,angle=270}
\vskip-5mm
\protect\caption{LMA $\pdod \ofwh$ vs $\w / \wk$ for $\htilde = h / \delno \zo = 0$, $0.2$, $0.5$, $1$, $2.5$ and $5$ (from top to bottom).}
\label{fig:7}
\end{figure}

In figure~\ref{fig:7} the $\htilde$-dependence of the resultant LMA spectrum is shown (for the same fields as in figure~\ref{fig:5} for the LMA(RPA)).  For $\htilde \lesssim 1$, the first effect of the field is to erode the zero-field Kondo resonance `on the spot' -- first diminishing and then splitting the resonance, but doing so largely under the envelope of the zero-field spectrum itself.  With further increasing field however, and again in contrast to the LMA(RPA) (figure~\ref{fig:5}), the split Kondo peaks broaden and diminish further in intensity; and their maxima move outside the zero-field spectral envelope, although the tail behaviour of the resonance is asymptotically common for all fields (as shown generally in \S 4.1).

The LMA spectra shown in figure~\ref{fig:7} agree rather well with those obtained from recent NRG calculations \cite{ref:costi} (see figure~3 therein).  In particular the spectral characteristics  outlined above are as found in the NRG calculations, and the qualitative similarity of figure~\ref{fig:7} and figure~3 of reference~\cite{ref:costi} is self-evident.  Detailed comparison may be made by noting that the LMA spectra at $\w = 0$ for a given value of $\htilde = h / \delno \zo$ (given explicitly by equation~(\ref{eq:scd0})), are in very good quantitative agreement with the $\w = 0$ NRG spectra at the same value of the ratio $H / \tk$ employed in reference~\cite{ref:costi}.  The resultant comparison as a function of frequency is not of course quantitatively perfect; but it is certainly rather good, and the LMA is to our knowledge the only theoretical approach that bears such comparison to the NRG data (see also figure~\ref{fig:walter} below).

With increasing field the split Kondo resonance becomes increasingly broad/diffuse (see figure~\ref{fig:7} and figure~\ref{fig:8} below), and we now comment on the large-$\htilde$ dependence of the peak maxima in the LMA $D \ofwh$ (or equivalently in the $D_\sigma \ofwh$).  This we find numerically to be of the form $|\wpeaks| \sim h \ln \htilde$ for  $\htilde = h / \delno \zo \gg 1$.  Hence $|\wpeaks|$ exceeds the value of $2h$ (the Zeeman splitting energy), generally regarded as the maximum spectral shift: see \eg \cite{ref:mw,ref:costi}.  For the physical reasons outlined in \S 3, it is we believe clear that the maximum shift in the transverse spin polarization propagator -- which directly probes spin-flip dynamics -- will not exceed an energy on the order of the free spin-$\frac{1}{2}$ Zeeman splitting $2h$.  We do not however know of a convincing reason, physical or otherwise, as to why this should also hold for the {\em single-particle} spectrum in the Kondo scaling regime (note that the NRG calculations of reference~\cite{ref:costi}, which report $|\wpeaks| < 2h$, are limited to comparatively small fields of $H / \tk \lesssim 5$ or so).  This issue should however be resolvable by NRG calculations, possibly necessitating use of the DM-NRG technique introduced recently by Hofstetter \cite{ref:hof}.  Indeed we add that preliminary calculations on the AIM in the strong coupling scaling regime, using the DM-NRG method, confirm spectral shifts in excess of $2h$ for sufficiently large fields \cite{ref:hofpriv,ref:pzpriv}; as seen directly in figures~\ref{fig:walter}, \ref{fig:8} below.

In the opposite limit of $h \ra 0$, an exact result for the spectral maximum $\wpeaks (h) = -\sigma |\wpeaks (h)|$ of the $\sigma$-spin spectrum $D_\sigma \ofwh$ may be obtained using Fermi liquid theory, as discussed in the  Appendix.  This leads to

\be
\label{eq:shift1}
|\wpeaks| = \frac{\rw(0)}{1 + \beta \delno \zo^2}h
\ee

\noindent where $\beta = \lim_{h \ra 0}[\lim_{\w \ra 0} (\sigtsi \ofwh / \w^2)]$.  In the trivial non-interacting limit where $\rw(0) = 1$ (and $\beta = 0$), this recovers correctly $|\wpeaks (h)| = h$ ($= \frac{1}{2}\gmb H$).  In the strong coupling Kondo limit by contrast, $\rw (0) = 2$ and the exact $\beta = 1 / (2 \delno \zo^2)$ \cite{ref:hewson}; hence

\be
\label{eq:shift2}
|\wpeaks (h)| = \frac{4}{3}~ h.
\ee

Equation~(\ref{eq:shift2}), valid for $\htilde \ll 1$, is exact for the Kondo model (albeit somewhat unexpected, conventional lore suggesting $|\wpeaks (h)| = 2h$).  It is not however recovered by the LMA for $D_\sigma \ofwh$ which instead yields $|\wpeaks(h)| = h$ as $\tilde{h} \ra 0$; reflecting the fact that while $\rw (0)=2$ arises correctly within the LMA \cite{ref:ld}, the resultant $\beta$ is readily shown to be twice the exact value.  Neither does equation~(\ref{eq:shift2}) appear to be recovered by the spinon approximation to the single-particle spectrum \cite{ref:mw} (discussed further in \S 5.3 below), the field dependence of $|\wpeaks (h)| / 2h$ being shown in figure~2 (top) of reference~\cite{ref:mw}.  For the lowest field considered there ($\gmb H / T_0 \equiv 2h/\wk \sim 1$), $|\wpeaks (h)| / 2h$ is close to but slightly less than $\frac{2}{3}$, and appears to be diminishing further with decreasing field (although its $h \ra 0$ asymptote has not to our knowledge been determined).  NRG calculations of the Kondo model were originally reported \cite{ref:costi} to yield $|\wpeaks(h)| \simeq 2h$ for $\htilde \ll 1$, but we understand that subsequent reanalysis of the low-field data \cite{ref:costipriv} is now consistent with the exact result equation~(\ref{eq:shift2}).

\begin{figure}
\centering\epsfig{file=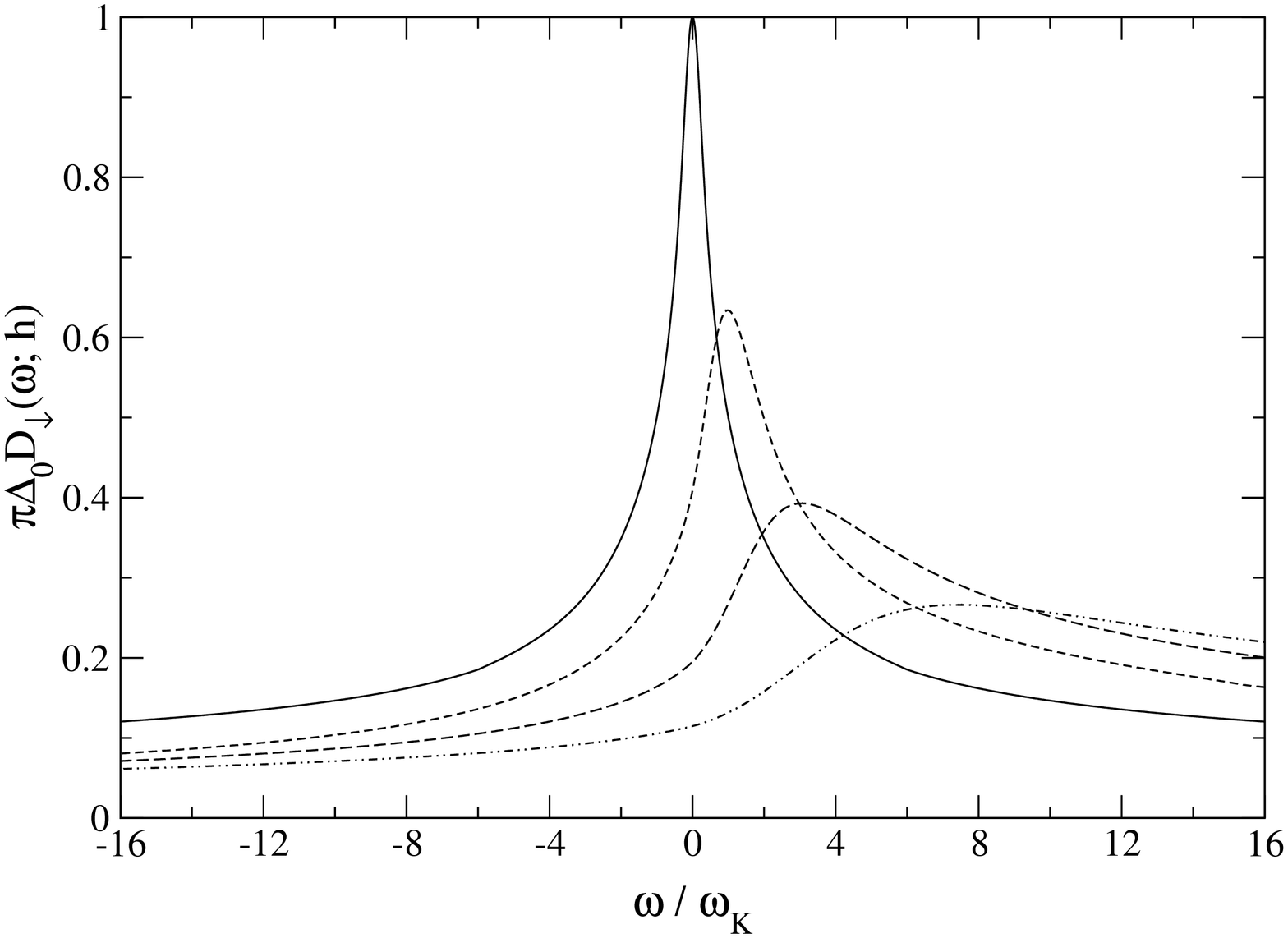,width=100mm,angle=270}
\vskip-5mm
\protect\caption{LMA $\pdod_\da \ofwh$ vs $\w / \wk$ for $\htilde = 1$ (short dash), $2.5$ (long dash), $5$ (point dash); and $\pdod \ofwo$ (solid line).}
\label{fig:8.5}
\end{figure}

Finally, recall from \S 2 that for $h=0$, $D \ofwo$ given from equation~(\ref{eq:absum}) is independent of spin $\sigma$, and is equivalently the zero-field $\sigma$-spin spectrum.  That the present LMA does not recover correctly the above low-field spectral shift of $D_\sigma \ofwh$, stems from the fact that the individual LMA $D_\sigma \ofwh $ as $h \ra 0$ do not separately coincide with $D\ofwo$ for arbitary $\w$ (although the differences in general are rather minor, and $D_\sigma \ofoh = D \ofoh$ for all $h$ and either $\sigma$).  This in turn is a natural consequence of the fact that (\S 2), for any $h \ne 0$, one or other MF state is picked out according to $\sgn (h)$; and as a result the peak maxima in the individual $D_\sigma \ofwh$ for small $\htilde$ are not captured correctly.  Note that the above remarks apply solely to the individual $D_\sigma \ofwh$, since $D \ofwh$ for all $\w$ evolves continuously in $h$ to its $h = 0$ limit $D \ofwo$ (see \S 2).  It is in fact possible to recover the correct low-field shift in $D_\sigma \ofwh$ within an LMA framework; but discussion of this would take us too far afield, and the present LMA does appear otherwise to account rather well for the overall spectral behaviour of the individual $D_\sigma \ofwh$.  These are illustrated in figure~\ref{fig:8.5} for $\htilde = 1$, $2.5$ and $5$, and exhibit the same behaviour as the NRG results shown in figure~2 of reference~\cite{ref:costi}.  In addition, the LMA for $\pdod_\da \ofwh$ is compared explicitly in figure~\ref{fig:walter} with results arising from the DM-NRG approach \cite{ref:hof} for a field $h / \wk = 25$ (kindly provided by W. Hofstetter \cite{ref:hofpriv}, and obtained for the AIM at a strong coupling $\tilde{U} = 8$).  The LMA is seen to account well for the DM-NRG data.  Note in particular that both approaches clearly yield peak maxima $|\wpeaks|$ that are in excess of the Zeeman splitting $2h$; and that, on the low-frequency side in particular, the slow logarithmic tails in $\pdod_\da \ofwh$ (given by the first term on the right hand side of equation~(\ref{eq:sclargewd})) are clearly evident.

\begin{figure}
\centering\epsfig{file=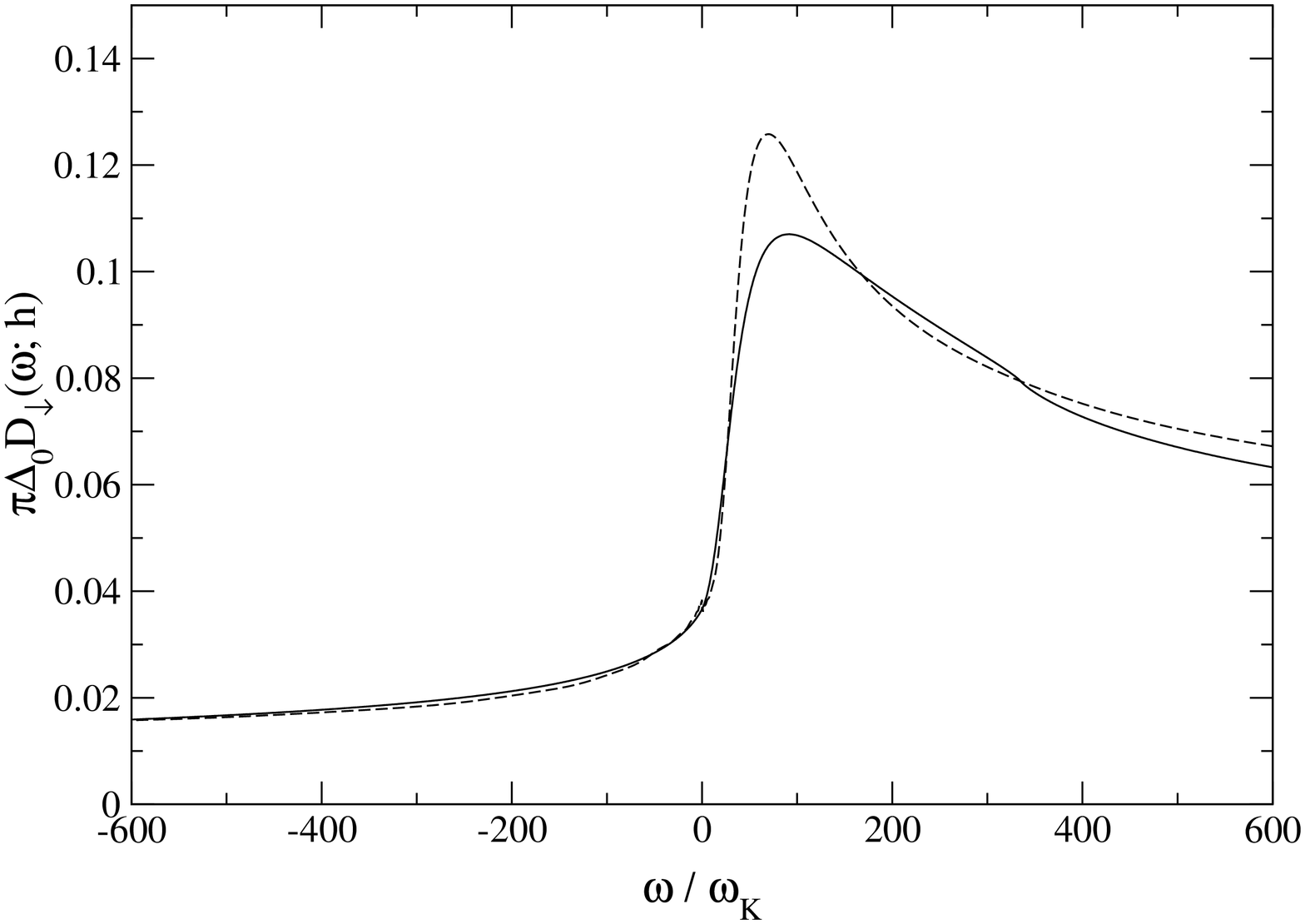,width=100mm,angle=270}
\vskip-5mm
\protect\caption{$\pdod_\da \ofwh$ vs $\w / \wk$ for $h / \wk = 25$.  Solid line: LMA.  Dashed line: result from DM-NRG calculations \cite{ref:hofpriv}.}
\label{fig:walter}
\end{figure}

\subsection{Spinon approximation.}

Here we compare the LMA results of the preceding section to the spinon approximation (SA) \cite{ref:mw}, in which the $h$-dependent single-particle spectrum is approximated by the density of states for spinon excitations obtained via the Bethe ansatz; and make further observations on the SA itself, based upon the results presented in reference~\cite{ref:mw}.  We denote the SA to the single-particle spectrum by $\rho_\sigma \ofwh$ (in the notation of reference~\cite{ref:mw}); with $\rho_\sigma \ofwh \propto D_\sigma \ofwh$.

To compare directly the LMA and SA we require the full proportionality, $\rho_\sigma \ofwh = C D_\sigma \ofwh$ with $C$ an $\w$- and $h$-independent constant.  This is clearly given by $C  = \rho_\sigma (0;0) / D_\sigma(0;0) = \rho_\sigma (0;0) \pi \delno$ (using $\pdod_\sigma (0;0) = 1$). It is however known \cite{ref:ko} that the zero-field spinon spectrum $\rho_\sigma (\w;0)$ is a pure Lorentzian, with a HWHM denoted by $T_0$ in reference~\cite{ref:mw} (such that $T_0 \equiv \wk$ in the notation of the present work).  Hence $\rho_\sigma(0;0)  = 1 / \pi T_0$, and $C = \delno / T_0$.  $\rho \ofwh = \sum\nolimits_\sigma \rho_\sigma \ofwh$ is thus related to $D \ofwh$ ($=\frac{1}{2}\sum\nolimits_\sigma D_\sigma \ofwh$) by

\be
\label{eq:rhotod}
\rho \ofwh 2h = \frac{4}{\pi} \frac{h}{\wk} \pdod \ofwh .
\ee

\begin{figure}
\centering\epsfig{file=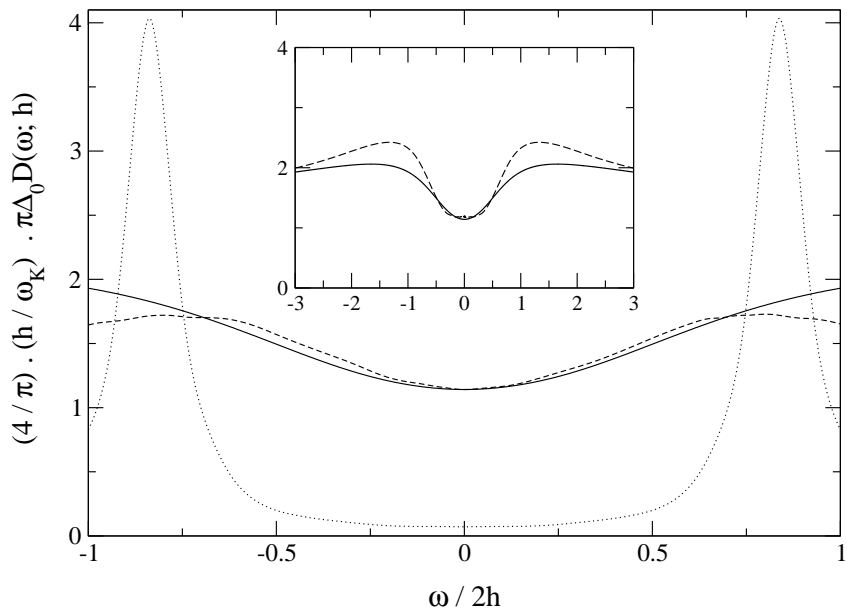,width=100mm,angle=270}
\vskip-5mm
\protect\caption{Spinon approximation to the single-particle spectrum \cite{ref:mw}, $\rho \ofwh 2h$ vs $\w / 2h$ (dotted line), compared to the corresponding LMA result $[4 h / \pi \wk] \pdod \ofwh$ (solid line); for a field $h / \wk = 24$.  The experimental data of reference~\cite{ref:qdots} is also shown (dashed line); see text (\S 5.3) for full discussion.  Inset: LMA spectrum on an expanded frequency scale (solid line), compared to $[4h/\pi \wk] \pdod \ofwh$ vs $\w / 2h$ obtained via the DM-NRG method \cite{ref:hof,ref:hofpriv} for $h / \wk = 25$ (dashed line).}
\label{fig:8}
\end{figure}

In figure~1 (bottom) of reference~\cite{ref:mw}, and for a field $2h / T_0 = 48$ ($\equiv 2h / \wk$), $\rho\ofwh 2h$ vs $\w / 2h$ is shown in the interval $|\w| / 2h < 1$ (to which frequency range the SA is restricted in practice \cite{ref:mw}).  In figure~\ref{fig:8} this SA result is compared to that arising from the LMA (via equation~(\ref{eq:rhotod})).  The two clearly differ very significantly for all $\w$.

To gain some insight into the above, it is natural to consider the functional form of the SA $\rho_\ua \ofwh$ (such that $\rho \ofwh = \rho_\ua \ofwh + \rho_\ua(-\w;h)$ by p-h symmetry).  In figure~1 (top) of reference~\cite{ref:mw}, $\rho_\ua \ofwh 2h$ vs $\w / 2h$ is shown, again for $h / T_0 = 24$.  Although not remarked upon in reference~\cite{ref:mw}, $\rho_\ua \ofwh$ appears to be a Lorentzian, given by

\be
\label{eq:salorentz}
\rho_\ua \ofwh = \frac{\frac{1}{2}\Delta E (h) \pi^{-1}  }{[ \w + E_{\rm max}(h)]^2 + [\frac{1}{2} \Delta E (h)]^2  }
\ee

\noindent where (using the notation of reference~\cite{ref:mw}) $E_{\rm max}(h)$ and $\Delta E (h)$ denote respectively the peak position and FWHM.  The numerical validity of equation~(\ref{eq:salorentz}) may be confirmed by direct transcription of the data given in figure~1 of reference~\cite{ref:mw} for $h / T_0 = 24$; the Lorentzian form equation~(\ref{eq:salorentz}) fits the data highly accurately (the correlation coefficient being 0.99995).

That $\rho_\sigma \ofwh$ should be a Lorentzian is not perhaps surprising, it being well known \cite{ref:ko} that the zero-field $\rho_\sigma (\w;0)$ is a pure Lorentzian; and although verified directly only for $h / T_0 = 24$, we naturally assume $\rho_\sigma \ofwh$ to be a Lorenztian for general $h / T_0$.  Granted this, the SA to the $\w = 0$ single-particle spectrum $D \ofoh$ ($\equiv D_\sigma \ofoh$) then follows directly from equations~(\ref{eq:rhotod}, \ref{eq:salorentz}) as:

\be
\label{eq:sastatics}
\pdod \ofoh \simeq \frac{T_0}{2h} \frac{(\Delta E(h) / 4h)} {\left[ \frac{E_{\rm max}(h)}{2h}\right]^2 + \left[ \frac{\Delta E (h)}{4h}\right]^2}.\qquad {\rm (SA)}
\ee

%hardref
\noindent The $h$-dependence of $D\ofoh$ is however known exactly (see equations~(\ref{eq:scmag}, 4.15)), enabling the validity of the SA to be ascertained from the $h$-dependence of $E_{\rm max}(h)$ and $\Delta E (h)$ given in reference~\cite{ref:mw}.  Defining $h_0 = h / T_0$, their asymptotic behaviours are known analytically for $h_0 \gg 1$ in particular \cite{ref:mw}: $E_{\rm max} (h) / 2h \sim 1 - {\cal O}([\ln(h_0)]^{-1})$ and $\Delta E (h) / 2h \sim {\cal O}([\ln^2(h_0)]^{-1})$.  Hence from equation~(\ref{eq:sastatics}), the SA to $\pdod \ofoh$ for large fields $h_0 \gg 1$ is $\pdod \ofoh \sim [h_0 \ln^2 (h_0)]^{-1}$.  This is not however the exact asymptotic behaviour, which from equation~(\ref{eq:scd0sh}) is in contrast given by $\pdod \ofoh \sim [\ln^2(h_0)]^{-1}$ (using $h_0 \propto \htilde = h / \delno \zo$).  And this in turn underlies the disparity evident in figure~\ref{fig:8} between the SA and LMA results for $D \ofoh$ in particular (for fields of this order the LMA is very close to the exact $D\ofoh$, see \eg figure~\ref{fig:4}).

The deficiencies of the SA are apparent even for $h=0$, where the resultant Lorentzian spectrum \cite{ref:ko} is simplistic, failing to recover the logarithmic tails known to dominate the single-particle scaling spectrum (see \eg \cite{ref:dl}).  The approximation does not moreover appear to improve for $h \ne 0$, as evident from the discussion above.  In this regard it is perhaps salient to note that the large-field asymptotic behaviour $\sim [h_0 \ln^2(h_0)]^{-1}$ deduced above for the $\w = 0$ spinon spectrum, is in fact that of the exact excess impurity susceptibility $\chi_{\rm i} (h)$ \cite{ref:afl}.  But whatever its physical content, the SA does not in our view provide a qualitatively satisfactory description of the single particle spectrum.

\subsection{Experiment.}

The differential conductance of a quantum dot in the presence of a magnetic field, $G_{\rm c}(V;h)$ with $V$ the applied (drain-source) voltage, has recently been studied experimentally by Goldhaber-Gordon et al \cite{ref:qdots}.  Their data for an applied field $H = 7.5{\rm T}$ has been compared to the SA results \cite{ref:mw}, taking $\gmb H / \wk = 2h/\wk = 48$ and $|g| = 0.36$ (as suggested by ESR measurements on 2DEGs \cite{ref:dvkw}).  With this we likewise compare the experimental data \cite{ref:qdots} to the LMA results of \S 5.2.  The two steps involved are as follows.  First, considering $T=0$, we take $(e^2 / 2 \pi \hbar )^{-1} G_{\rm c} (V;h) \propto \pdod (\w = V;h)$.  This proportionality holds strictly for the linear differential conductance ($V = 0$), non-equilibium effects arising for $V \ne 0$ thereby being neglected in practice (as also in \eg reference~\cite{ref:mw}).  Second, the proportionality constant $\gamma$ is determined simply fom the experimental linear differential conductance, viz $\gamma = [(e^2 / 2 \pi \hbar)^{-1} G_{\rm c} \ofoh ] / (\pdod \ofoh)$ with $(e^2 / 2 \pi \hbar)^{-1} G_{\rm c} \ofoh$ taken directly from the data in reference~\cite{ref:qdots} and $D \ofoh$ the LMA spectrum at $\w = 0$.

The latter step requires an explanation, since the LMA spectrum is for $T=0$ while the experiment \cite{ref:qdots} is performed at $T = 90{\rm mK}$ (approximately twice the Kondo temperature).  The Kondo resonance is continuously destroyed by the separate effects of temperature (as controlled by the ratio $T/\wk$) and an applied magnetic field (controlled by $\gmb H / \wk$).  But if $\gmb H \gg T$ the latter effect overwhelms the former, and comparison to the $T=0$ limit is thus warranted (the validity of which argument is in fact consistent with the NRG results of reference \cite{ref:costi}, figure~3).  This is indeed the relevant case in the experiment \cite{ref:qdots} at $H = 7.5{\rm T}$ and $T = 90 {\rm mK}$, where (with $|g| = 0.36$ as above) $\gmb H / T \sim 20$.

The resultant comparison between the LMA and experiment \cite{ref:qdots} is shown in figure~\ref{fig:8} as a function of $\w / 2h$ (with $[4h/\pi\wk]\pdod \ofwh$ shown to enable comparison to the SA results \cite{ref:mw} also given in the figure).  The agreement with experiment is on the whole rather good, particularly for $|\w| \lesssim h$ where the $\w$ ($\equiv V$) dependence of the experimental $G_{\rm c}(V; h)$ appears to be quite well accounted for by the LMA.  Insofar as the modest differences between experiment and the LMA may be attributed to non-equilibrium effects, neither these nor (as argued above) thermal effects would thus appear to dominate the experimental observations.  In this regard our interpretation of experiment differs significantly from that of reference~\cite{ref:mw}; which is natural given the marked disparity (figure~\ref{fig:8}) between experiment and the SA results \cite{ref:mw}.  The LMA is in turn in rather good agreement with results obtained from the DM-NRG approach \cite{ref:hof,ref:hofpriv}, particularly for $|\w| \lesssim h$.  This is seen from the inset to figure~\ref{fig:8} where the above LMA results are compared to DM-NRG data for $2h / \wk = 50$ (see also figure~\ref{fig:walter}).

%The inset to figure~\ref{fig:8} also compares the above LMA results to those obtained from the DM-NRG approach \cite{ref:hof} for $2h / \wk = 50$ \cite{ref:hofpriv}.
%(kindly provided by W. Hofstetter \cite{ref:hofpriv}).  The LMA is seen to account rather well for the DM-NRG data, particularly for $|\w| \lesssim h$.  Note further, in relation to the spectral shifts discussed in \S 5.2, that the DM-NRG results clearly show peak maxima that are in excess of the Zeeman splitting $2h$.

\section{Summary.}

%hard ref
The subject of this paper has been single-particle dynamics of the symmetric Anderson impurity model in the presence of a magnetic field. While topical, the problem certainly poses a significant theoretical challenge. The non-crossing approximation [33-35,24] for example, which despite its inability to recover Fermi liquid behaviour at low-energies has been used to considerable effect for $H=0$ (see \eg \cite{ref:hewson}), fails quite dramatically for $H \neq 0$; producing spurious spectral peaks at the Fermi level \cite{ref:mwl} that are symptomatic \cite{ref:mw} of its origins as a large-$N$ theory. Likewise equation of motion approaches \cite{ref:mwl} simply lack the low-energy Kondo scale,  modified perturbation theory \cite{ref:ts} is confined to weak coupling, and the spinon approximation \cite{ref:mw} suffers from the qualitative limitations discussed in \S 5.3.

The local moment approach [16-19] developed here transcends many of the limitations of previous theoretical approaches. All energy scales, field and interaction strengths are handled by it, leading thereby to a rather comprehensive description of the problem that recovers in particular the important strong coupling, Kondo scaling regime. Here, as for the zero-field case \cite{ref:dl}, the LMA appears to pass the acid test of comparison to benchmark numerical calculations provided by the NRG \cite{ref:costi,ref:hof,ref:hofpriv}; as well as yielding rather good agreement with experiments on quantum dots \cite{ref:qdots}. Its strengths stem in part from its inherent simplicity and physical transparency, together with the fact that it is not confined \eg  to problems that are ubiquitously Fermi liquid-like on low-energy scales \cite{ref:L2,ref:bulla1}. As such, we believe it provides a powerful tool for further investigation of a wide spectrum of quantum impurity physics, and related lattice-based models within the framework of dynamical mean-field theory \cite{ref:vollhardt,ref:gkkr}.

\ack
We are grateful to R. Bulla, W. Hofstetter, T. Pruschke and R. Zitzler for numerous helpful discussions, with particular thanks for Dr. Hofstetter for kindly providing the DM-NRG results shown in figures~\ref{fig:walter}, \ref{fig:8}.  We are also grateful to the EPSRC, the Leverhulme Trust and the British Council for financial support.

\renewcommand{\appendix}{%
	\setcounter{section}{1}
	\setcounter{equation}{0}
	\renewcommand{\thesection}{\Alph{section}}
	\renewcommand{\theequation}{\Alph{section}.\arabic{equation}}
}

\appendix

\section*{Appendix}

Using microscopic Fermi liquid theory, we obtain two exact results for the Kondo limit of the AIM; specifically for the field dependence of the quasiparticle weight ($\zh / \zo$) and the asymptotic low-field behaviour of the spectral shifts ($\wpeaks$) in $D_\sigma \ofwh$.

The basic underlying equations are as follows.  The (excess) impurity magnetization $M_\rmi (h)$ follows directly from the Friedel sum rule (see \eg \cite{ref:hewson}); and for the symmetric AIM is given generally by

\be
\label{eq:mi}
M_\rmi (h) = \frac{\gmb}{\pi} \tan ^{-1} \left[ (h - \sigma \sigtsr \ofoh) / \delno \right]
\ee

\noindent (where $\sigma \sigtsr \ofoh$ is independent of spin, see equation~(\ref{eq:sigsym})).  From this the corresponding impurity susceptibility follows, $\chi_\rmi (h) = \frac{1}{2} \gmb (\partial M_\rmi / \partial h)$.   The Wilson ratio $\rw (h) = c \chi_\rmi (h) / \gamma_\rmi (h)$, with $\gamma_\rmi (h)$ the linear specific heat coefficient ($c = [2\pi k_{\rm B}]^2 / [3(\gmb)^2]$); it is given by

\be
\label{eq:rw}
\rw (h) = \zh [ 1 - \sigma(\partial \sigtsr \ofoh / \partial h)].
\ee

\noindent And using (\ref{eq:mi}, \ref{eq:rw}), $\chi_\rmi (h) $ may be expressed as

\be
\label{eq:chi}
\chi_\rmi (h) = \frac{(\gmb)^2}{2} D \ofoh \frac{\rw (h)}{\zh}
\ee

\noindent with $D \ofoh \equiv D_\sigma \ofoh$ the spectrum at the Fermi level $\w = 0$.  Finally, in the Kondo regime of vanishing charge susceptibility, we will employ the Ward identity \cite{ref:hewson}

\be
\label{eq:fullward}
1 + \sigma\left( \frac{\partial \sigtsr \ofoh}{\partial h} \right) = 2 \left( \frac{\partial \sigtsr \ofwh}{\partial \w}\right)_{\w = 0}
\ee

\noindent (which is not specific to the symmetric model).

Consider first the Wilson ratio.  Since $\zh$ is given by equation~(\ref{eq:zhdef}), equations~(\ref{eq:rw}, \ref{eq:fullward}) yield directly $\rw (h) = 2 ~ \forall ~ h$.  For $h=0$, this result is well known since the classic work of ${\rm Nozi\grave{e}res}$ \cite{ref:noz}.  Its validity for all $h$ was first suggested by Wiegmann and Finkelstein \cite{ref:wf} and later proven using the Bethe ansatz \cite{ref:tw}.  It is not of course specific to the symmetric model: the generalization of (\ref{eq:rw}) for the asymmetric AIM is readily shown to be $\rw (h) = \sum\nolimits_\sigma D_\sigma \ofoh [ 1 -\sigma (\partial \sigtsr \ofoh / \partial h)] / \sum\nolimits_\sigma D_\sigma \ofoh [ 1 - (\partial \sigtsr \ofwh / \partial \w)_{\w = 0}]$; combined with (\ref{eq:fullward}), $\rw(h) = 2 ~ \forall~h$ again results.

We now consider $\zh / \zo$.  From (\ref{eq:rw}), using $\rw (h) = 2$,

\be
\label{eq:zh1}
\frac{\zh}{\zo} = \frac{[1 - \sigma(\partial \sigtsr \ofoh / \partial h)_{h = 0}]}{[1 - \sigma ( \partial \sigtsr \ofoh / \partial h )]} .
\ee

\noindent With $\sigtsr \ofoh$ given from (\ref{eq:mi}),

\be
\label{eq:zh2}
1 - \sigma ( \partial \sigtsr \ofoh / \partial h) = \sec^2 \left[ \frac{\pi M_\rmi (h)}{\gmb} \right]~ \frac{2 \pi \delno}{(\gmb)^2}~ \chi_\rmi (h)
\ee

\noindent and hence

\be
\label{eq:zh3}
\frac{\zh}{\zo} = \frac{\cos^2 \left[ \frac{\pi M_\rmi (h)}{\gmb} \right]}{\tilde{\chi}_\rmi (h)}.
\ee

\begin{figure}
\centering\epsfig{file=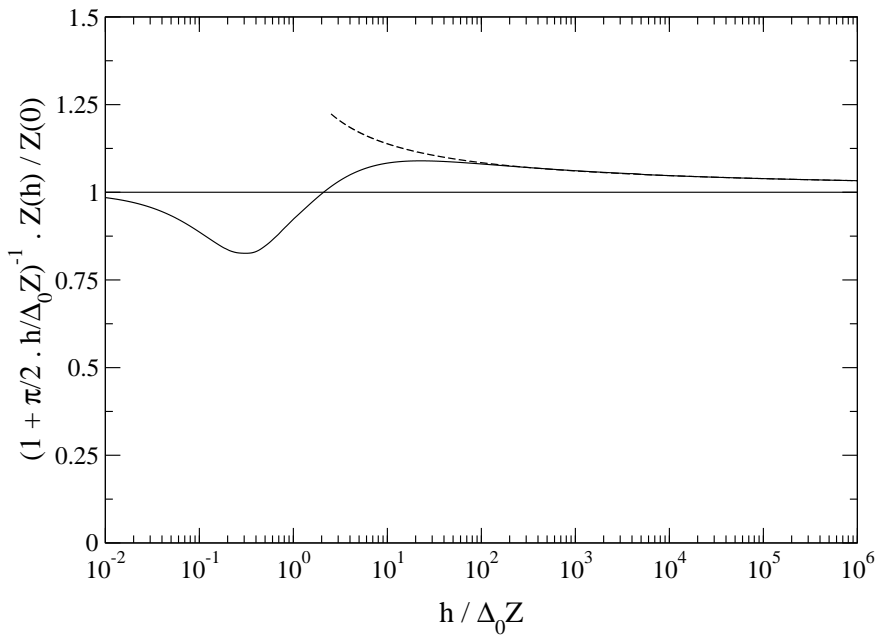,width=100mm,angle=270}
\vskip-5mm
\protect\caption{Exact field dependence of the quasiparticle weight, $[\zh / \zo] / (1 + \frac{\pi}{2}\htilde)$ vs $\htilde = h / \delno \zo$ (solid line);  the large-$\htilde$ asymptote (equation~(\ref{eq:zhcorrections})) is also shown (dashed line).}
\label{fig:zcomp}
\end{figure}

\noindent Here $\tilde{\chi}_\rmi (h) = \chi_\rmi (h) / \chi_\rmi(0)$; and from (\ref{eq:chi}) using $\rw(0) = 2$ and $\pdod (0;0) = 1$, $\chi_\rmi (0) = (\gmb)^2 / (\pi \delno \zo)$ ($\equiv (\gmb)^2 / 4 k T_{\rm L}$ where $k T_{\rm L} = \frac{\pi}{4}\delno \zo$).  Equation~(\ref{eq:zh3}) is exact, and $M_\rmi (h)$ (and hence $\chi_\rmi (h)$) is likewise known exactly from the BA solution of the Kondo model \cite{ref:afl}.  The $\htilde = h / \delno \zo$ dependence of $\zh / \zo$ may thus be found explicitly, and is shown in figure~\ref{fig:zcomp} where $[\zh / \zo] / [1 + \frac{\pi}{2}\htilde]$ is plotted; this being the ratio of the exact $\zh / \zo$ to that arising within the LMA (see equation~(\ref{eq:zhz0})).  The latter is thereby seen to concur well with the exact result, deviation from which is typically $\lesssim 10\%$ save for $\htilde \sim {\cal O}(1)$ where it is  slightly more.

The large-$\htilde$ behaviour of $\zh / \zo$ bears note, it being known from the  BA solution that \cite{ref:afl}

\be
\label{eq:lhmag}
M_\rmi (h) ~~ {_{\htilde \gg 1}\atop^{\sim}}~~ \frac{\gmb}{2} \left[  1 - \frac{1}{2 \ln (b\htilde)} - \frac{\ln \ln (b\htilde)}{4[\ln(b\htilde)]^2} + {\cal O}\left([\ln \htilde]^{-3}\right) \right]
\ee

\noindent (where $b = 4 \sqrt{e / \pi}$).  The LMA for $M_\rmi (h)$ (discussed in \cite{ref:ld}) recovers correctly the leading logarithmic approach to saturation ($[2 \ln (\htilde)]^{-1}$), but not the subleading $\ln\ln(\htilde)/[\ln(\htilde)]^2$ corrections.  It is in fact the latter that generate the leading logarithmic corrections to $\zh / \zo \sim \frac{\pi}{2} \htilde$ for $\htilde \gg 1$; which using (\ref{eq:lhmag}) and (\ref{eq:zh3}) are given by

\be
\label{eq:zhcorrections}
\frac{\zh}{\zo}  ~~ {_{\htilde \gg 1}\atop^{\sim}}~~ \frac{\pi}{2} \htilde \left[ 1 + \frac{1}{2\ln (b\htilde)} \right].
\ee

\noindent The asymptotic behaviour (\ref{eq:zhcorrections}) is also shown in figure~\ref{fig:zcomp} (dashed line).

Finally, we obtain an exact result for the asymptotic low-field behaviour of the spectral shifts.  $G_\sigma \ofwh$ itself is given by equation~(\ref{eq:gsig}), with $\sigts \ofwh $ the exact self-energy; and we consider explicitly the wide-band AIM for which $\Delta (\w) = -\rmi~\sgn (\w) \delno$ (this does not of course impose any restrictions on results for the Kondo regime).  $D_\sigma \ofwh = -\pi^{-1}\sgn(\w)\im G_\sigma \ofwh$ follows from equation~(\ref{eq:gsig}), and the spectral maximum therein arises for $\w = \wpeaks(h)$ such that $(\partial D_\sigma \ofwh / \partial \w)_{\wpeaks} = 0$; to which equation we seek the asymptotic low-field solution with $\wpeaks \propto -\sigma h$, say $\wpeaks = - \gamma \sigma h$.  This is merely a matter of algebra: one needs simply a standard low-frequency expansion of the self-energy, $\sigtsr \ofwh - \sigtsr \ofoh \sim - [ \zh^{-1} - 1]\w$ and $\sigtsi \ofwh \sim \beta (h) \w^2$; and to recognize from (\ref{eq:rw}) that $\lim_{h \ra 0} [1 - \frac{\sigma}{h}\sigtsr \ofoh]\zh \equiv \rw(0)$.  With $\beta \equiv \beta(h=0)$, the general result for the AIM is thereby found to be

\be
\label{eq:lhpeak}
|\wpeaks| = \frac{\rw(0)}{[1 + \beta \delno \zo^2]}h
\ee

\noindent as discussed in \S 5.2.

\vskip+5mm

\end{document}